\documentclass[prl,preprintnumbers,amsmath,amssymb,twocolumn,longbibliography]{revtex4-1}
\usepackage{amssymb}
\usepackage{amsmath}
\usepackage{graphics}
\usepackage{graphicx}
\usepackage{subfigure}
\usepackage{dcolumn}
\usepackage{bm}
\usepackage{hyperref}
\usepackage{enumerate}
\usepackage{color}

\begin{document}

\preprint{APS/123-QED}

\title{Linking Dispersion and Stirring in Randomly Braiding Flows}

\author{Daniel R. Lester}
\email[]{daniel.lester@rmit.edu.au}
\affiliation{School of Engineering, RMIT University, Melbourne, Australia}
\author{Michael G. Trefry}
\affiliation{Independent Researcher, Perth, Australia}
\author{Guy Metcalfe}
\affiliation{Swinburne University of Technology, Melbourne, Australia}
\date{\today}

\begin{abstract}
Many random flows, including 2D unsteady and stagnation-free 3D steady flows, exhibit non-trivial braiding of pathlines as they evolve in time or space. We show that these random flows belong to a pathline braiding \emph{universality class} that quantitatively links dispersion and chaotic stirring, meaning that the Lyapunov exponent can be estimated from the purely advective transverse dispersivity.  We verify this quantitative link for both unsteady 2D and steady 3D random flows.  This result uncovers a deep connection between transport and mixing over a broad class of random flows.
\end{abstract}


\maketitle

Most steady and unsteady flows exhibit complex Lagrangian kinematics, whereby pathlines intertwine in a chaotic manner in space and time~\cite{Arnold:1965aa,Henon:1966aa}. Such Lagrangian chaos or chaotic adection~\cite{Aref:1984aa,Ottino:1989aa} arises from stretching and folding of fluid elements, leading to highly striated material distributions that profoundly augment many fluid-borne phenomena~\cite{Aref:2017aa,Metcalfe:2012aa,Speetjens:2021aa}, including solute transport \cite{Jones:1994aa,Lester:2014ab}, mixing~\cite{Cerbelli:2017aa,Fereday:2004aa}, chemical reactions~\cite{Tel:2005aa,Neufeld:2009aa}, biological activity~\cite{Tel:2000aa,Karolyi:2000aa}, particle transport~\cite{Ouellette:2008aa,Haller:2008aa} and heat transfer~\cite{Lester:2009aa,Baskan:2015aa}.  Complex stirring arises because the kinematic equation 
 $\dot{\mathbf{x}}(t)=\mathbf{v}(\mathbf{x},t)$
is rich enough to admit chaotic dynamics if the velocity field $\mathbf{v}(\mathbf{x},t)$ possesses sufficient spatio-temporal degrees of freedom ($\text{dof}\geqslant 3$)~\cite{Poincare:1893aa} and constraints such as symmetries in Stokes flow~\cite{Haller:1998ab} or helicity in Darcy flow~\cite{Metcalfe:2023aa} are not imposed.

\begin{figure}
\begin{centering}
\begin{tabular}{c c}
\multicolumn{2}{c}{\includegraphics[width=0.8\columnwidth]{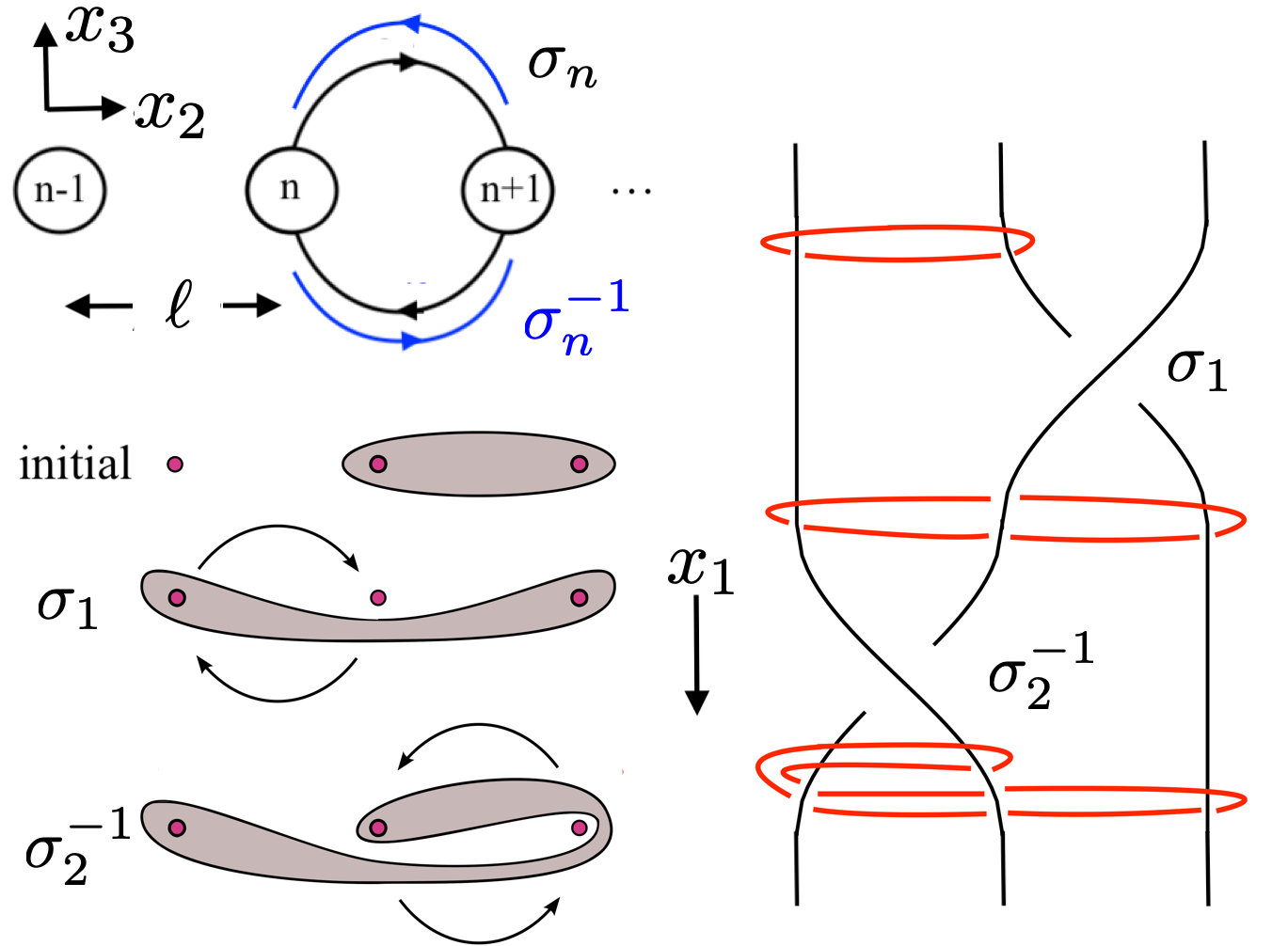}}\\
\multicolumn{2}{c}{(a)}\\
\includegraphics[width=0.47\columnwidth]{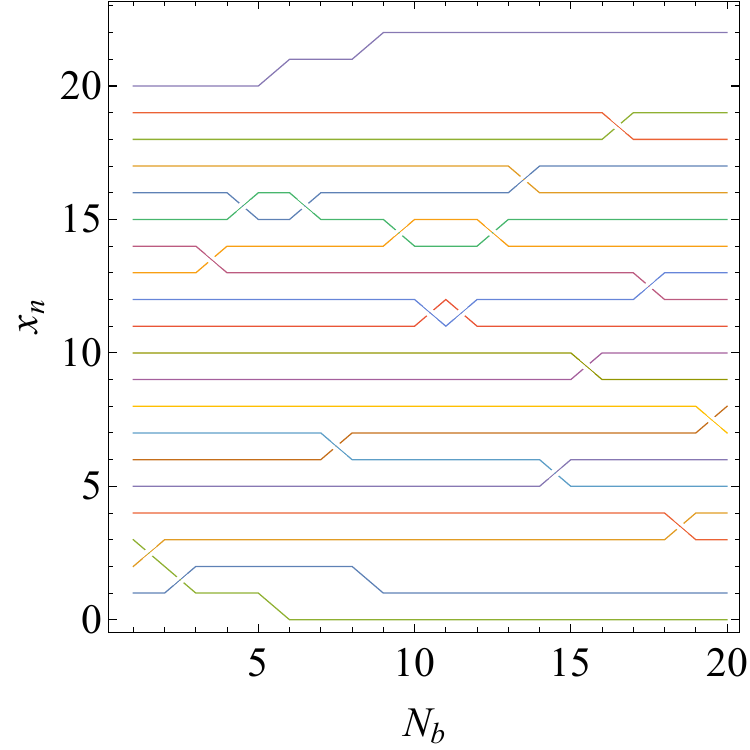}&
\includegraphics[width=0.48\columnwidth]{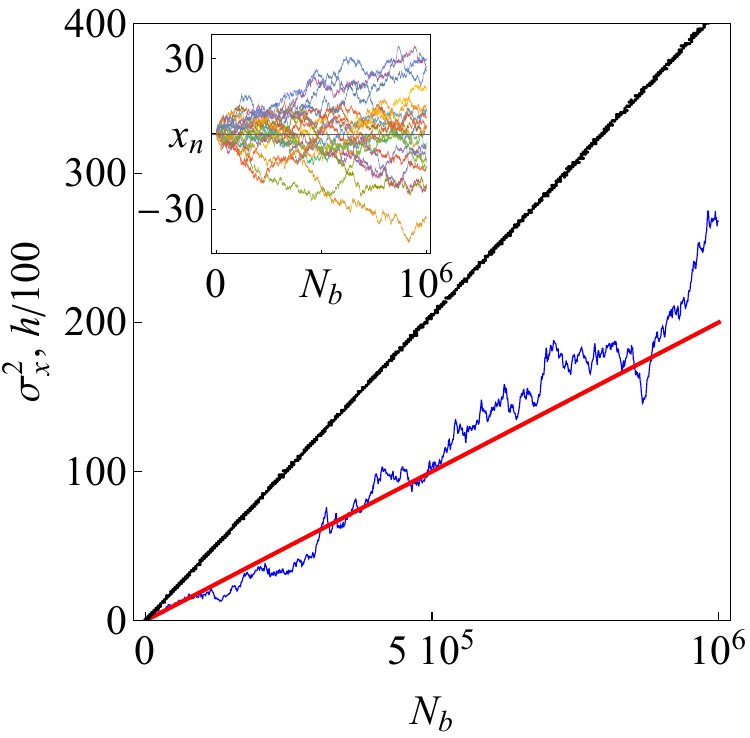}\\
 (b) & (c)
\end{tabular}
\end{centering}
\caption{
(a) Top left: Schematic of the 1D streamline model including streamlines $n-1$, $n$, $n+1$ in the transverse $x_2-x_3$ plane and clockwise (black) $\sigma_n$, anti-clockwise (blue) $\sigma_n^{-1}$ braid generators . Bottom left: stretching of material elements due to braiding motions (adapted from \cite{braidbook}) that evolve with the longitudinal $x_1$ direction. Bottom right: braid diagram depicting stretching of material elements (red) around streamlines (black). 
(b) Braid diagram for $N_p=20$ streamlines over $N_b=20$ braid actions in the $x_1$ direction, leading to non-trivial braiding and transverse dispersion. 
(c) Growth of topological entropy $h$ (black line) and transverse variance $\sigma_{x_2}^2$ (blue line) with $N_b$, in agreement with (\ref{eqn:variance}) (red line). Inset: Brownian motion of streamlines with increasing $N_b$.}
\label{fig:braid}
\end{figure}

Unsteady 2D flows and stagnation-free steady 3D flows may be topologically unified by considering the former as the latter with unit velocity in time~\cite{Bajer:1994aa}. For these $\text{dof}=3$ flows, the intertwining of pathlines may be formalised as a non-trivial \emph{braiding} process~\cite{Boyland:2000aa}, and a mathematical framework~\cite{Handel:1985aa} and associated tools~\cite{Moussafir:2006aa,Bestvina:1995aa} have been developed to quantify the topological complexity of the braiding motions. As pathlines are invariant 1D objects in a 3D continuum, non-trivial pathline braiding stretches and folds the interstitial continuum fluid in a chaotic manner (Fig.~\ref{fig:braid}a)~\cite{Boyland:2000aa}. As the topological complexity of the braiding motions approximates that of the fluid flow~\cite{Thiffeault:2010aa}, there exist deep connections between braiding complexity and the rate of chaotic stirring~\cite{Thiffeault:2005aa}.

In addition to stirring, pathline braiding in unbounded random flows is also strongly linked to advective transverse particle dispersion as random pathlines must typically cross one another (with respect to an arbitrary reference direction) to separate without bound. Although there exist particular flows that do not braid but can disperse (e.g. hyperbolic flows) and vice-versa (e.g. confined braiding flows), these specific cases arise in either bounded or deterministic flows. For unbounded random flows, strong links have been observed between non-trivial braiding and particle dispersion in 2D turbulence~\cite{Francois:2015aa} and ferrofluids~\cite{Skjeltorp:1999aa}, while many studies~\cite{Pedersen:1996aa} have uncovered links between particle dispersion and chaotic stirring. Conversely, steady 3D flows with zero helicity density~\cite{Moffatt:1969aa,Moreau:1961aa}
$\mathcal{H}(\mathbf{x})\equiv\mathbf{v}(\mathbf{x})\cdot\nabla\times\mathbf{v}(\mathbf{x})$
are integrable~\cite{Arnold:1965aa} (non-chaotic) and admit a pair of streamfunctions $\psi_1(\mathbf{x})$, $\psi_2(\mathbf{x})$~\cite{Yoshida:2009aa,Lester:2021aa} that act as Euler potentials as
   $\mathbf{v}(\mathbf{x})=\nabla\psi_1(\mathbf{x})\times\nabla\psi_2(\mathbf{x})$.
These potentials prohibit pathline braiding as streamlines are confined to level sets of $\psi_1(\mathbf{x})$, $\psi_2(\mathbf{x})$ and have been proved~\cite{Lester:2023aa} to also prevent transverse particle dispersion. While these observations illustrate the strong correlations between chaotic stirring and particle dispersion in randomly braiding flows, lacking is a mechanistic understanding of the underlying link between these phenomena.

We examine the link between particle dispersion and braiding via simple random braiding models and show that there exists a pathline braiding \emph{universality class} to which all 3 dof unbounded random flows belong. This reveals the deep connection between dispersion and chaotic stirring in randomly braiding 3 dof flows; moreover, this uncovers a quantitative link between the Lyapunov exponent $\lambda_\infty$ and advective transverse dispersivity $D_T$. We validate the link for two example flows: an unsteady random 2D flow and an steady 3D random flow. The latter case is related to a  proof of the ubiquity of chaotic advection in heterogeneous Darcy flow~\cite{Lester:2024ab}.
 
To elucidate the link between pathline braiding, chaotic advection and transverse dispersion, we begin with a particularly simple streamline model that, nonetheless, captures the essential features of a mean longitudinal steady chaotic 3D flow: namely braiding of streamlines as they propagate longitudinally in the $x_1$ direction. The unsteady 2D analog is a 2D flow with pathlines that braid in time $t$ rather than $x_1$. As shown in Fig.~\ref{fig:braid}a (top left), the 1D streamline model consists of a 1D array of $N_p$ streamlines in $\mathbb{R}^3$ which propagate with uniform velocity in the $x_1$ direction and repeat periodically with uniform spacing $\Delta x_2=\ell$ in the $x_2$ direction, where $\ell$ is the velocity correlation length.

At integer multiples of longitudinal distance $x_1=\Delta_L$, one pair of neighbouring streamlines randomly exchange position via the clockwise or counter-clockwise rotations shown in Fig.~\ref{fig:braid}a, which respectively are labelled by the \emph{braid generators} $\sigma_n$ and $\sigma_n^{-1}$ for the streamline pair $(n, n+1)$, with $n\in 1:N_p-1$. The sequence of $N_b$ braid generators forms the braid word $\mathbf{b}=\sigma_{n_1}^{\pm1}\sigma_{n_2}^{\pm1}\dots\sigma_{n_{N_b}}^{\pm1}$ that completely defines the braiding of streamlines. Due to periodicity the set of $N_p$ streamlines undertake unbounded random walks along the $x_2$ coordinate as they propagate longitudinally. Fig.~\ref{fig:braid}b shows the braid diagram for $N_p=20$ streamlines undertaking $N_b=20$ random braid actions, illustrating braiding actions and transverse dispersion of streamlines.

\begin{figure}
\begin{centering}
\includegraphics[width=0.8\columnwidth]{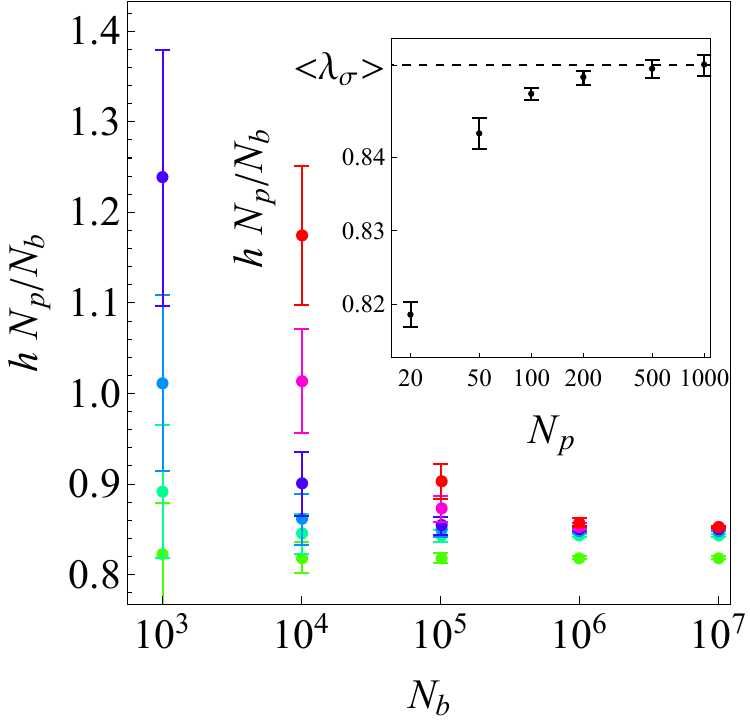}
\end{centering}
\caption{Convergence of scaled mean topological entropy $N_p/N_b \langle h\rangle$ with braid number $N_b$ for different streamline numbers $N_p$ (green $50$, light blue $100$, purple $200$, pink $500$, red $1000$). Dots indicate mean value, error bars indicate $\pm$ standard error. Inset: convergence of scaled mean topological entropy $h N_p/N_b$ for $N_b=10^7$ with streamline number $N_p$ to random braid entropy $\langle\lambda_\sigma\rangle$.}
\label{fig:entropy}
\end{figure}

Fig.~\ref{fig:braid}a shows that streamline braiding can lead to rapid stretching of material elements. If the braiding is non-trivial (pseudo-Anosov)~\cite{Boyland:2000aa}, this stretching grows exponentially with the number $N_b$ of braiding actions, the growth rate of which is quantified by the topological braid entropy $h$~\cite{braidbook} which converges to the topological entropy $\tilde{h}$ of the underlying fluid flow in the limit of large $N_p$~\cite{Thiffeault:2010aa}. Hence for statistically stationary random flows, the braid entropy $h$ converges to the Lyapunov exponent $\hat{\lambda}_\infty$ of the associated chaotic flow in the limit of large $N_p$ and $N_b$~\cite{Thiffeault:2005aa}. To construct a random braiding sequence we chose $N_b$ braid generators with equal probability from the set $\{\sigma_1,\sigma_1^{-1},\dots,\sigma_{N_p},\sigma_{N_p}^{-1}\}$, and calculate the topological entropy $h$ via the \emph{braidlab} package~\cite{braidlab}. Streamline braiding in this model is governed by the correlation length $\ell$, and in \cite{suppmat} we show that additional streamlines (i.e. $\Delta x_2<\ell$) only contribute trivial braiding motions and so do not alter the braiding entropy $h$. 

Fig.~\ref{fig:braid}c shows the topological entropy $h$ grows linearly with $N_b$, indicating such random braiding generates exponential growth of material elements. As two of the $N_p$ streamlines make random jumps in the $x_2$ direction of $\pm\ell$ with each braiding event, the spatial variance $\sigma_{x_2}^2$ of streamlines grows linearly with $N_b$ as
\begin{equation}
\sigma_{x_2}^2(t)=\sigma_{x_2,0}^2(0)+2\ell^2\frac{N_b}{N_p}\equiv\sigma_{x_2,0}^2(0)+2D_T t,\label{eqn:variance}
\end{equation}
where $N_b=\langle v_1\rangle t/\Delta_L$, $D_T$ is the transverse dispersivity and $\langle v_1\rangle$ is the mean longitudinal velocity. As shown in Fig.~\ref{fig:braid}c, this predicted Brownian motion agrees well with model observations.

Fig.~\ref{fig:entropy} shows that for $10^3$ realisations of random braids with $N_p$ particles and $N_b$ braids, $N_b/N_p \langle h\rangle$ converges to the \emph{random braid entropy} $\langle\lambda_\sigma\rangle$
\begin{equation}
\lim_{N_p,N_b\rightarrow\infty}\frac{N_p}{N_b}\langle h\rangle\rightarrow\langle\lambda_\sigma\rangle\approx 0.8529\approx\frac{2}{3}\left(\frac{2\pi}{3}\right)^{1/3},
\end{equation}
which quantifies the topological braiding entropy of unbounded random 3 dof flows. Hence for this model the dimensionless Lyapunov exponent $\lambda_\infty\equiv\hat{\lambda}_\infty\ell/\langle v_1\rangle$ is linearly related to the transverse dispersivity $D_T$ as
\begin{equation}
\lambda_\infty=\frac{\langle\lambda_\sigma\rangle}{Pe_T}.
\label{eqn:1Dmodel}
\end{equation}
where $Pe_T\equiv D_T/(\langle v_1\rangle\ell)$ is the advective transverse P\'{e}clet number. This linear relationship is expected as the 1D streamline model generates transverse dispersion and chaotic advection in equal measure.

To extend this model to 2D, we consider a periodic square array of $N_p=N_x\times N_x$ streamlines in the $x_2-x_3$ plane with spacing $\Delta x_2=\Delta x_3 = \ell$ as shown in Fig.~\ref{fig:2Dbraid}a. Similar to the 1D model, rotations of pairs of adjacent streamlines are chosen at random, but streamline pairs may now be chosen in either direction, leading to linear variance growth
\begin{align}
    \sigma^2_{x_2}(t)=\sigma^2_{x_3}(t)=\sigma^2_0+\frac{2}{d}\ell^2\frac{N_b}{N_p}=\sigma^2_0+2D_T t,
    \label{eqn:2Dvar}
\end{align}
where the factor of $2/d$ (where $d$ is the dimension of the streamline array) arises as each rotation moves two streamlines by $\pm\ell$ in the $x_2/x_3$ direction $1/d$ of the time.

\begin{figure}
\begin{centering}
\begin{tabular}{c c}
\includegraphics[width=0.44\columnwidth]{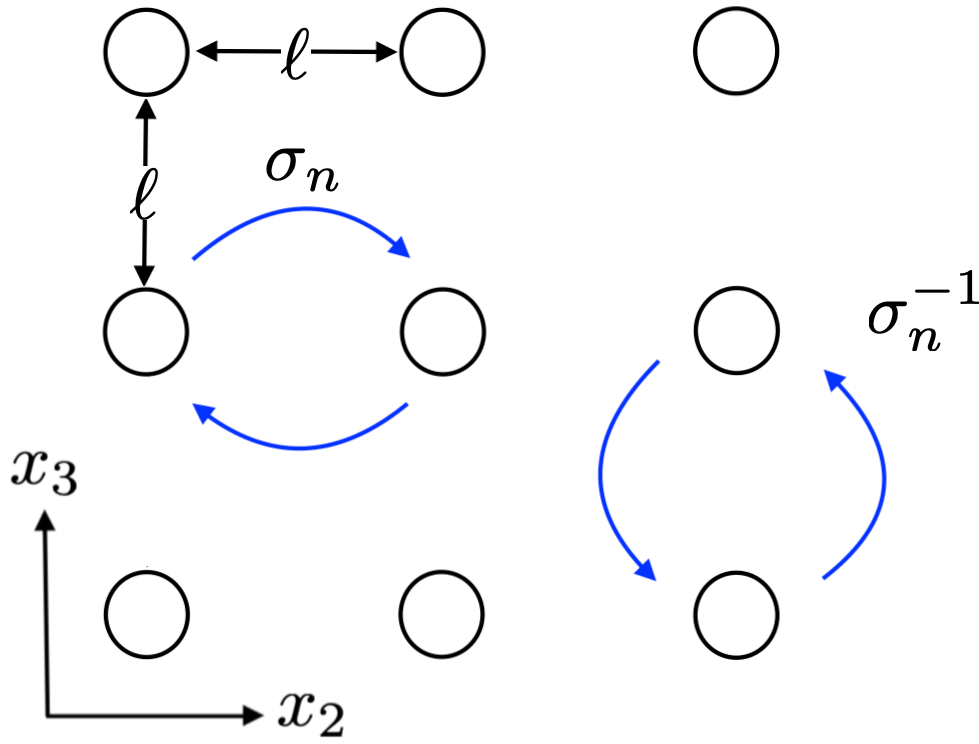}&
\includegraphics[width=0.54\columnwidth]{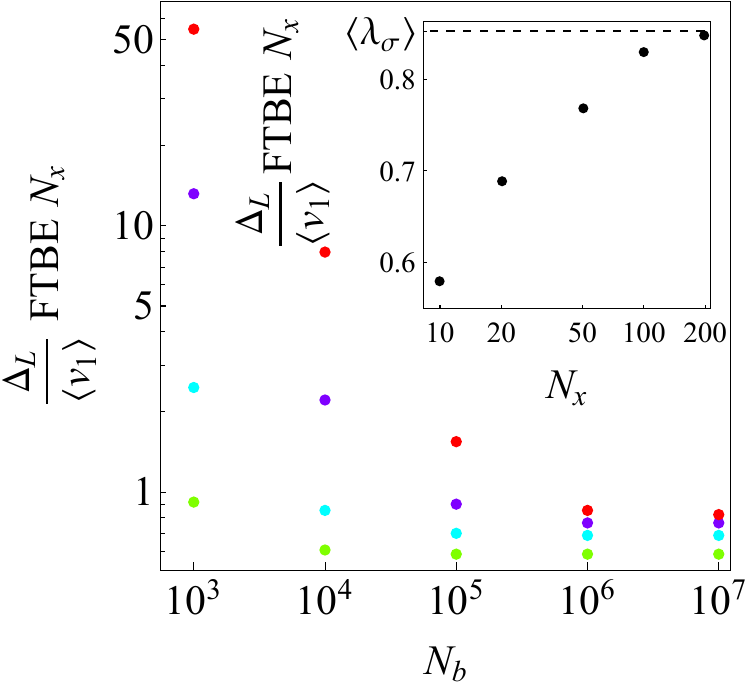}\\
(a) & (b)
\end{tabular}
\end{centering}
\caption{
(a) Schematic of 2D streamline model. (b) Convergence of scaled FTBE with $N_b$ for the 2D streamline model for different $N_x$ (green $N_x=10$, light blue $N_x=20$, purple $N_x=50$, pink $N_x=100$, red $N_x=200$). Inset: convergence of scaled FTBE for $N_b=10^7$ with $N_x$ to $\langle\lambda_\sigma\rangle$.}
\label{fig:2Dbraid}
\end{figure}

For the 2D streamline model, determination of the braid word from the sequence of braiding events is more complex. Hence the braiding topological complexity is quantified~\cite{suppmat} in \emph{braidlab} via the finite time braiding exponent (FTBE), an analogue of the finite time Lyapunov exponent which converges to $h$ as $\lim_{t\rightarrow\infty} h/t\rightarrow\lim_{t\rightarrow\infty}\text{FTBE}=\hat{\lambda}_\infty$~\cite{braidbook,Budisic:2015aa}. In ~\cite{suppmat} we establish convergence of the scaled FTBE and scaled $h$ to $\langle\lambda_\sigma \rangle$ with increasing $N_b$ for the 1D streamline model over $10^3$ realisations. As the FTBE is normalised by the advection time $t$ over which the braiding occurs, the FTBE has units of inverse time and so the prefactor $\Delta_L/\langle v_1\rangle$ renders the scaled FTBE dimensionless.

For each braid word, the corresponding streamlines are constructed (see Fig.~\ref{fig:2Dbraid}b) that resolve the clockwise and anti-clockwise rotations, and the \emph{databraid} routine is used to compute the FTBE from these streamlines. The topological entropy is also computed in the usual manner via the \emph{annbraid} routine in \emph{braidlab}. In the limit of large particle numbers $N_p$ and braid events $N_b$, the scaled FTBE and topological entropy converge to each other and the random braiding exponent $\langle\lambda_\sigma\rangle$.\\
For the 2D model, we construct 3D streamline trajectories (shown in  \cite{suppmat}) from a random braiding sequence and compute the FTBE. Fig.~\ref{fig:2Dbraid}b shows convergence of the scaled average FTBE over 10 realisations to the random braid entropy with increasing $N_b$, $N_p$ as $\lim_{N_b\rightarrow\infty}\lim_{N_p\rightarrow\infty}\Delta_L/\langle v_1\rangle\,\text{FTBE}\,N_x\rightarrow\langle\lambda_\sigma\rangle$.
\citet{Budisic:2015aa} show that in limit of long times and large particle numbers the FTBE converges to the Lyapunov exponent $\hat{\lambda}_\infty$, hence
\begin{equation}    \frac{\ell}{\langle v_1\rangle}\lim_{N_b\rightarrow\infty}\lim_{N_p\rightarrow\infty}\,\text{FTBE}=\lambda_\infty,
\end{equation}
and so the Lyapunov exponent is related to $\langle\lambda_\sigma\rangle$ as
\begin{equation}
    \lambda_\infty=\frac{\ell}{\langle v_1\rangle}\,\text{FTBE}=\langle\lambda_\sigma\rangle\frac{\ell}{\Delta_L}\frac{1}{N_x}.
\end{equation}
and using (\ref{eqn:2Dvar}) yields an expression consistent with (\ref{eqn:1Dmodel})
\begin{equation}
    \lambda_\infty^d
    =\langle\lambda_\sigma\rangle^d\left(\frac{\ell}{\Delta_L}\right)^{d-1}\frac{d}{Pe_T},\quad d=1,2.\label{eqn:model}
\end{equation}

In \cite{suppmat}, we extend the 2D streamline model to a 2D random walk model (Fig.~\ref{fig:2DRandom}a) which, following \cite{braidbook}, is comprised of $N_p$ random streamlines in the domain $\Omega:(x_1,x_2,x_3)\in[0,\infty)\times[0,L]\times[0,L]$ that make random jumps of magnitude $\Delta_T$ in the $x_2-x_3$ plane at integer multiples of the longitudinal distance $x_1=\Delta_L$. As shown in  Fig.~\ref{fig:2DRandom}b, the scaled FTBE for this system also converges to $\langle\lambda_\sigma\rangle$ for different jump sizes $\Delta_T$, and recovers~\cite{suppmat} the relationship (\ref{eqn:model}) between transverse dispersivity and Lyapunov exponent. 

\begin{figure}
\begin{centering}
\begin{tabular}{c c}
\includegraphics[width=0.35\columnwidth]{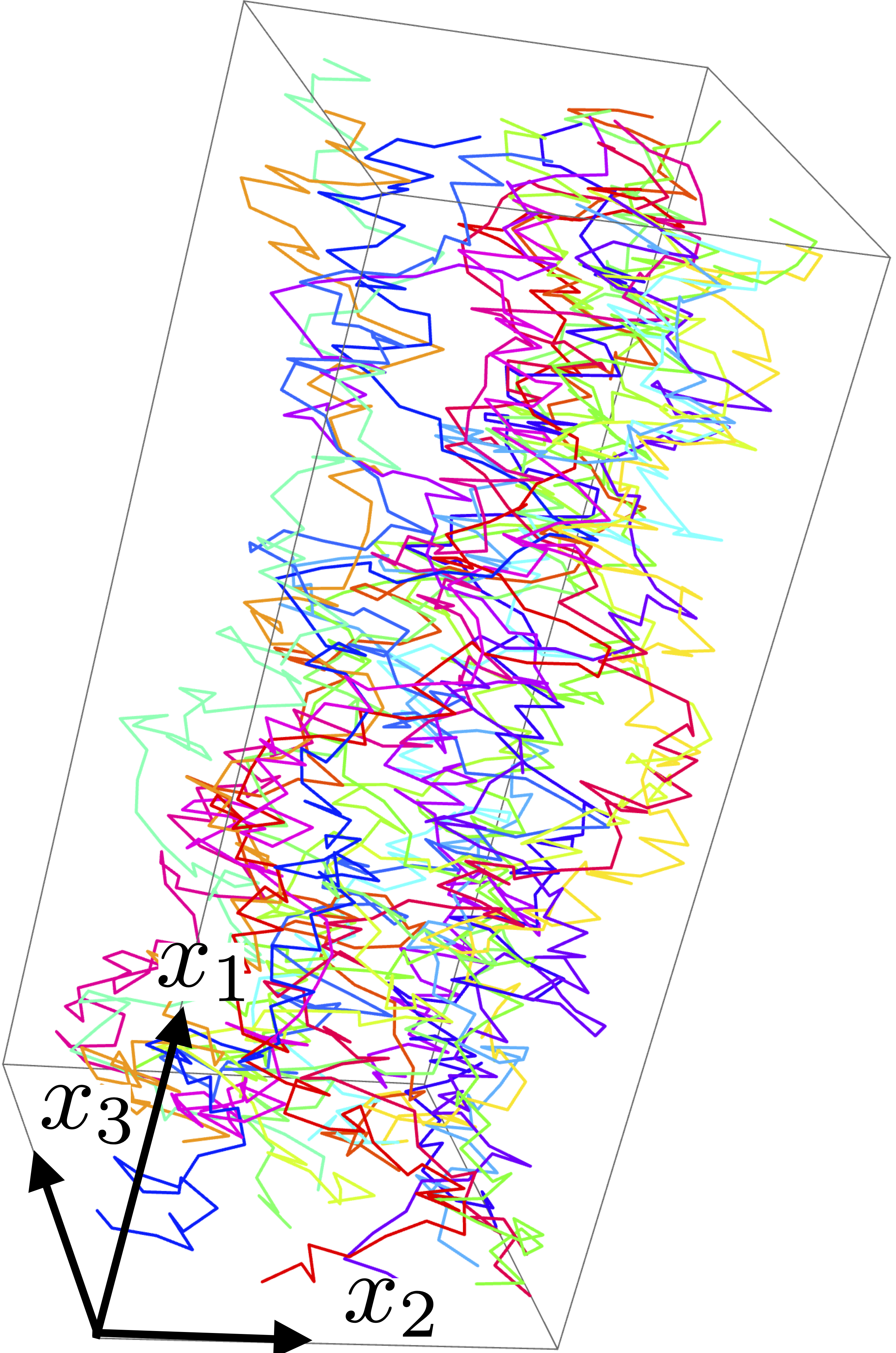}&
\includegraphics[width=0.63\columnwidth]{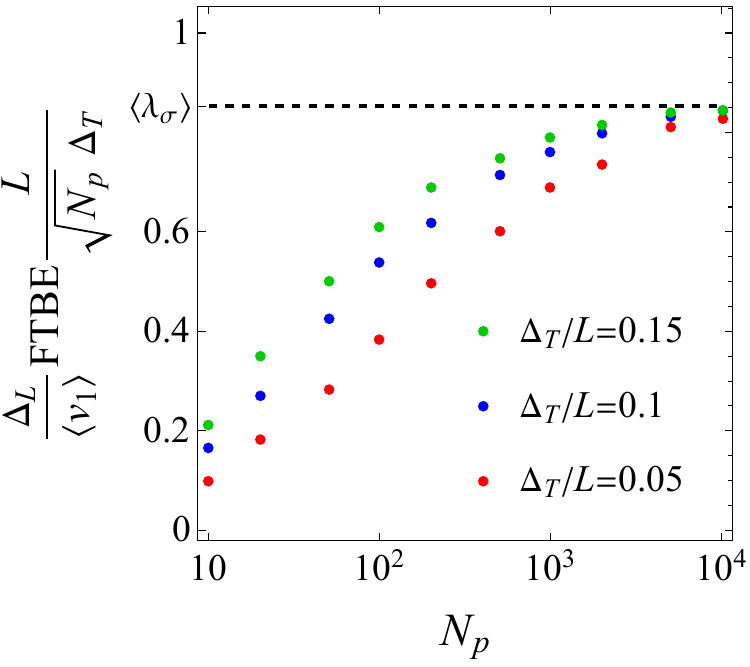}\\
(a) & (b)
\end{tabular}
\end{centering}
\caption{(a) Schematic of 2D random walk streamline model. (b) Convergence of scaled FTBE to $\langle\lambda_\sigma\rangle$ for $N_b=10^6$ with $N_p$ for different values of $\Delta_T/L$.}
\label{fig:2DRandom}
\end{figure}
 
The persistence of (\ref{eqn:model}) and $\langle\lambda_\sigma\rangle$ suggests that these diverse models all belong to the same \emph{universality class}~\cite{Odor:2004aa} associated with streamline braiding. This shows that chaotic advection and transverse dispersion are intimately linked in unbounded random 3 dof flows because they are both driven by non-trivial streamline braiding. 

\begin{figure}
\begin{centering}
\begin{tabular}{c c}
\includegraphics[width=0.40\columnwidth]{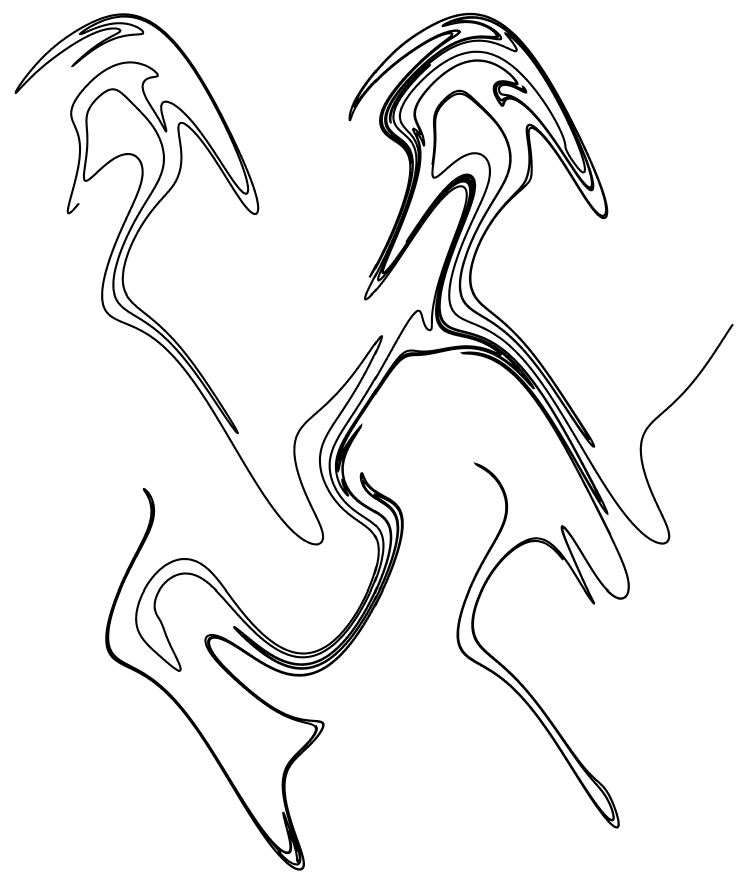}&
\includegraphics[width=0.58\columnwidth]{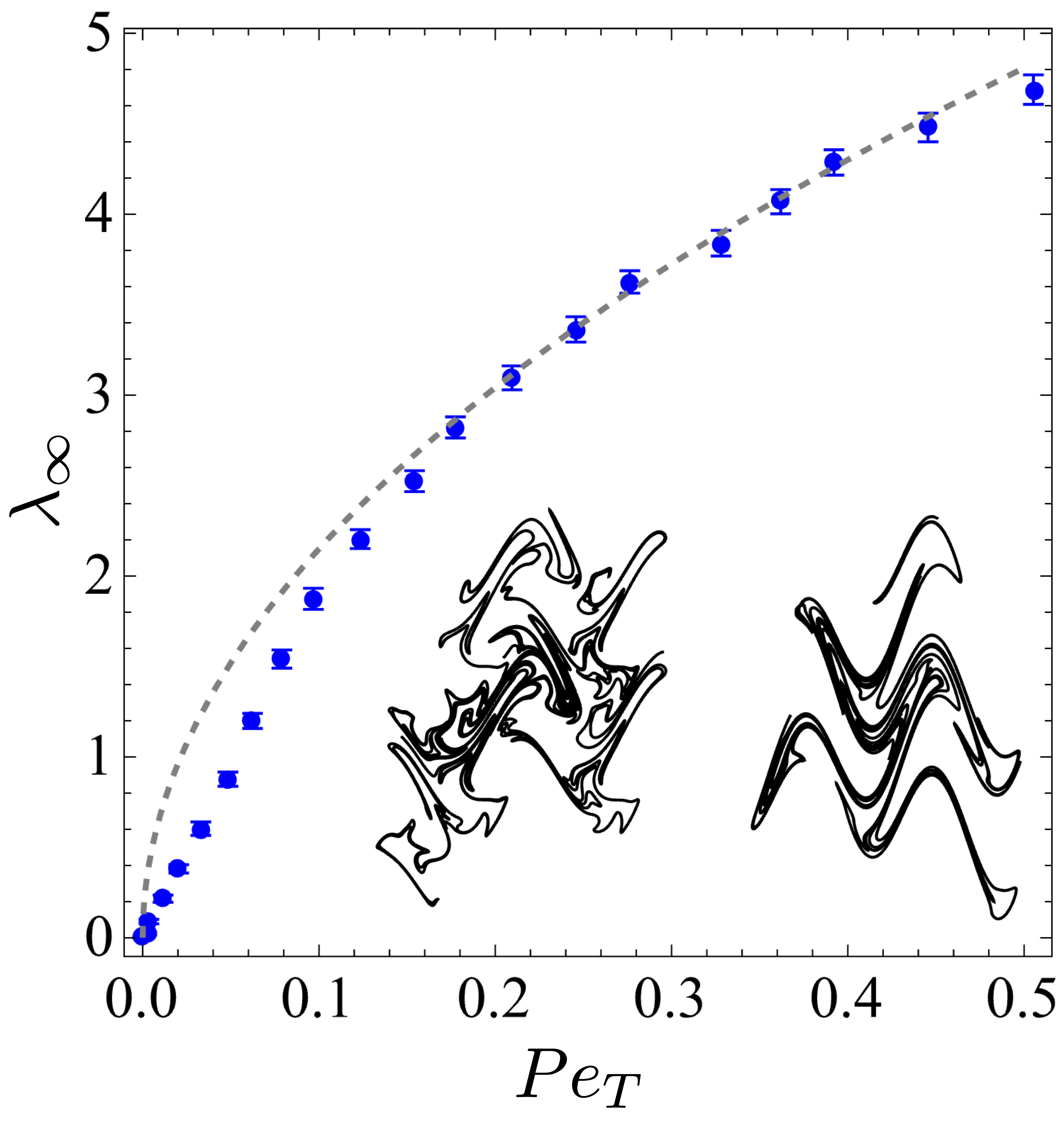}\\
(a) & (b)
\end{tabular}
\end{centering}
\caption{(a) Dispersion, stretching and folding of a material line in a realisation of the TPRSF for $T=0.5$. (b) Correlation between Lyapunov exponent $\lambda_\infty$ and transverse P\'{e}clet number $Pe_T$ from (\ref{eqn:model}) (dashed line) and numerical simulations (blue dots). Insets show typical material lines for $T=0.25$ (left) and $T=0.5$ (right).}
\label{fig:sineflow}
\end{figure}

To examine the connection between dispersion, stirring and braiding in unsteady 2D flows, we consider the time-periodic random sine flow (TPRSF)~\cite{Alvarez:1998aa}
\begin{equation}
\mathbf{v}(\mathbf{x},t)=
\begin{cases}
(0,\sin(2\pi x_1+\zeta))\quad 0\leqslant \text{mod}(t,T)<T/2,\\
(\sin(2\pi x_2+\zeta),0)\quad T/2\leqslant \text{mod}(t,T)<T,
\end{cases}\label{eqn:sineflow}
\end{equation}
where the phase angle $\zeta\in[0,2\pi]$ are random uniformly distributed variables and the flow period $T\in[0,1]$ controls the transition from non-chaotic steady flow at $T=0$ to chaotic dynamics for $T>0$. Fig.~\ref{fig:sineflow}a shows that under the non-trivial braiding action of the flow, tracer particles disperse outwards in the $x_1$ and $x_2$ directions, and material lines are stretched and folded in a complex manner. Fig.~\ref{fig:sineflow}b shows the Lyapunov exponent $\lambda_\infty$ calculated for $10^4$ independent tracer particles (see \cite{suppmat} for details) and the transverse P\'{e}clet number $Pe_T$ calculated for $10^6$ tracer particles (see \cite{suppmat} for details) for various values of $T\in[0,1]$. The grey dashed line in Fig.~\ref{fig:sineflow}b shows the predicted Lyapunov exponent from (\ref{eqn:model}), which is in excellent agreement with the numerical data.

To examine this connection for steady 3D flow, we consider heterogeneous Darcy flow governed by
\begin{equation}
    \mathbf{v}(\mathbf{x})=-\mathbf{K}(\mathbf{x})\cdot\nabla\phi(\mathbf{x}),\quad\nabla\cdot\mathbf{v}=0,\label{eqn:Darcy}
\end{equation}
in a triply-periodic unit cube (3-torus $\mathbb{T}^3$) $\mathcal{D}:\mathbf{x}\in[0,1]\times[0,1]\times[0,1]$ driven by a unit potential gradient $\nabla\bar{\phi}=\{-1,0,0\}$. We consider a heterogeneous porous medium with the following hydraulic conductivity field
\begin{equation}
\mathbf{K}(\mathbf{x})=k_0(\mathbf{x})\mathbf{I}+\delta(k_\delta(\mathbf{x})-k_0(\mathbf{x}))\,\hat{\mathbf{e}}_1\otimes\hat{\mathbf{e}}_1,\label{eqn:perturb}
\end{equation}
where $k_0(\mathbf{x})\neq k_\delta(\mathbf{x})$ and the perturbation parameter $\delta\in[0,1]$ quantifies the deviation of the conductivity tensor from isotropy, and the scalar log-conductivity fields $f(\mathbf{x})\equiv\ln k(\mathbf{x})$ are given by the random Fourier modes with correlation length $\ell=1/4$ and log-variance $\sigma^2_{\ln K}=4$, as detailed in \cite{suppmat}.
This is the simplest possible conductivity field that admits non-zero total helicity $H\equiv\int_{\mathcal{D}} d^3\mathbf{x}\,\mathcal{H}(\mathbf{x})$~\cite{Lester:2024ab}. Details regarding computation of the velocity field and streamlines, transverse dispersivity $D_T$ and Lyapunov exponent $\lambda_\infty$ are given in \cite{suppmat}.

\begin{figure}
\begin{centering}
\begin{tabular}{c c}
\includegraphics[width=0.48\columnwidth]{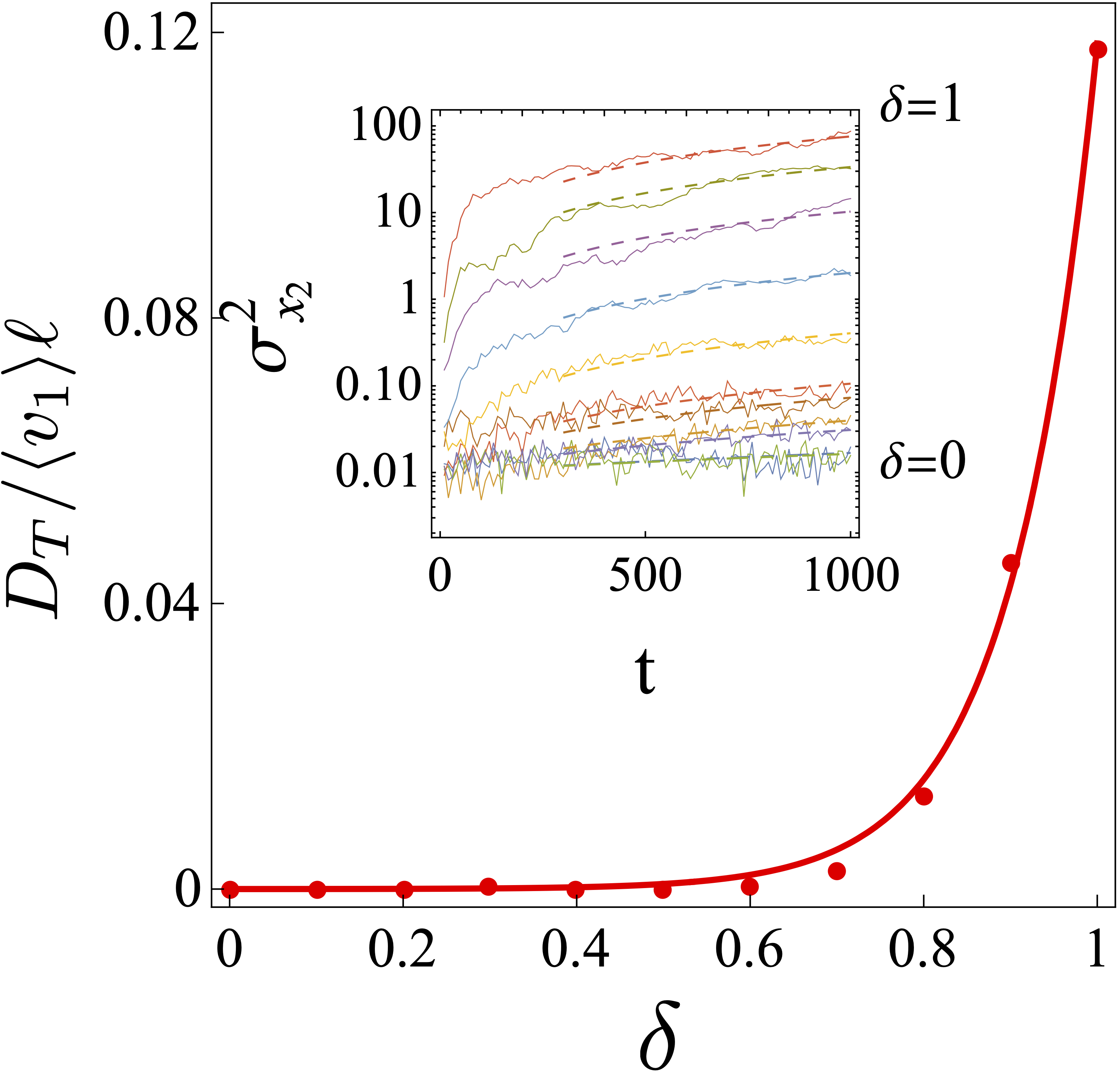}&
\includegraphics[width=0.46\columnwidth]{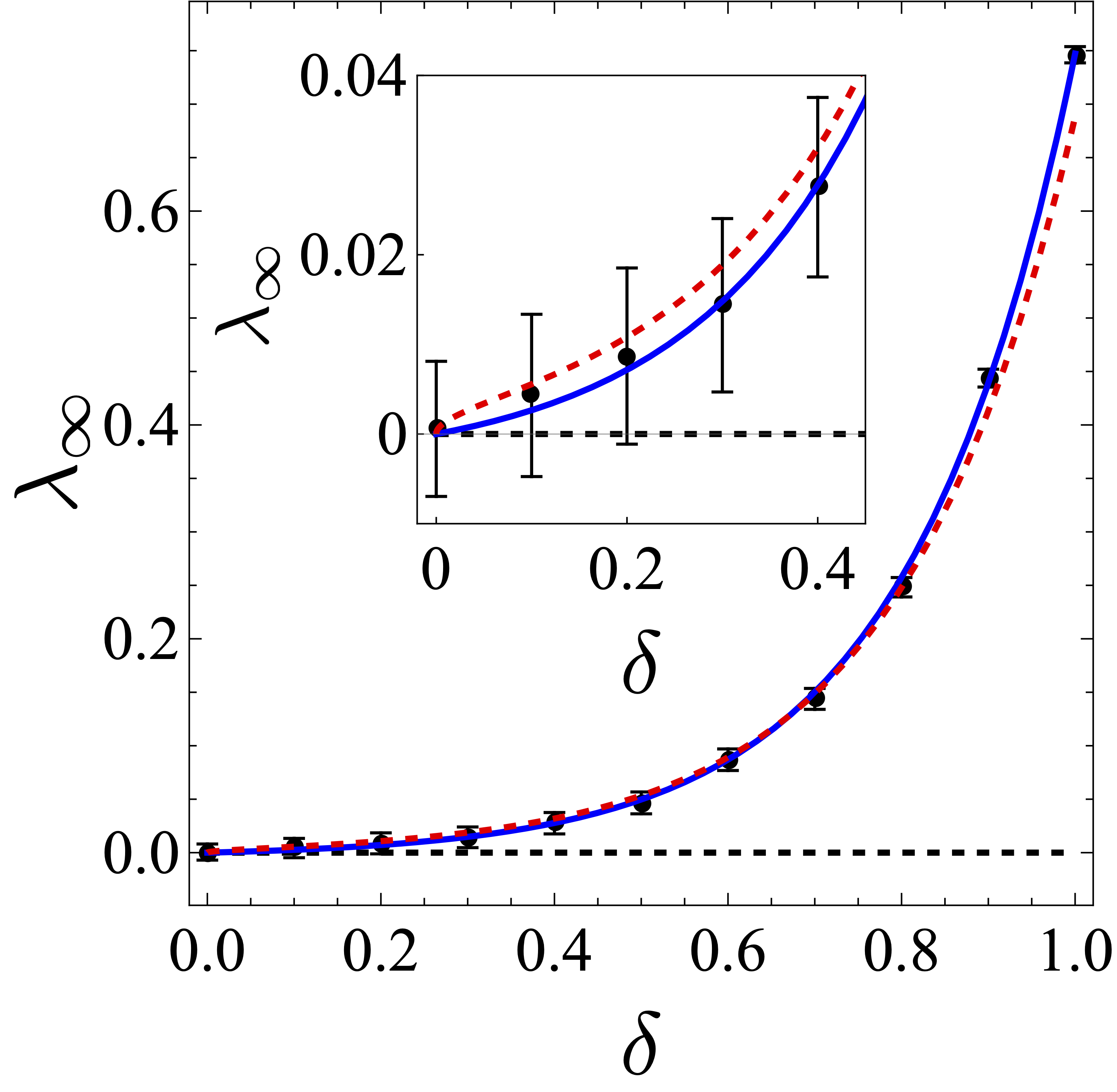}\\
(a) & (b)
\end{tabular}
\end{centering}
\caption{
(a) Growth of transverse P\'{e}clet number $Pe_T$ with $\delta$ from simulations (red points), and fitted exponential function (red curve). Inset shows temporal evolution of transverse variance. (b) Growth of Lyapunov exponent $\lambda_\infty$ with perturbation parameter $\delta$ from simulations (black points), and fitted exponential (blue curve). Error bars correspond to $\pm$ two standard deviations $\sigma_{\epsilon^\prime_{22}}$. (red dotted curve) Lyapunov exponent predicted from fitted exponential in (b) and (\ref{eqn:model}).}
\label{fig:delta}
\end{figure}

Fig.~\ref{fig:delta}a shows that the transverse P\'{e}clet number $Pe_T$ increases exponentially with $\delta$ as $Pe_T\approx 0.00381(e^{5.283\,\delta}-1)$, indicating transverse dispersion arises for weak perturbations away from isotropic Darcy flow. Fig.~\ref{fig:delta}b shows that the Lyapunov exponent $\lambda_\infty$ also increases exponentially with $\delta$ as $\lambda_\infty\approx 4.343\times 10^{-6}(e^{10.214\,\delta}-1)$, and is also non-zero for small $\delta>0$, indicating that chaotic advection occurs for weak perturbations away from heterogeneous isotropic Darcy flow. Fig.~\ref{fig:delta}b also shows that insertion of the fitted exponential for $Pe_T$ into (\ref{eqn:model}) (assuming $\Delta_L=\ell$ due to isotropy of the fields $k_0$, $k_\delta$) yields excellent agreement with the measured Lyapunov exponent.

These results clearly demonstrate a fundamental relationship between dispersion and chaotic stirring across a diverse range of random 3 dof flows.  The connection arises as both processes are generated by random pathline braiding. The existence of a pathline braiding universality class is demonstrated by the persistence of the universal braiding entropy $\langle\lambda_\sigma\rangle$ and the simple relationship (\ref{eqn:model}) linking these phenomena over a wide range of random flows. This link is especially important for steady 3D heterogeneous Darcy flow, as it forms the basis of a proof \cite{Lester:2024ab} of the ubiquity of chaotic advection in these flows. More broadly, the existence of a quantitative link between advective dispersion and chaotic stirring in unbounded random 3 dof flows provides insights into their fundamental kinematics and has implications for understanding, quantifying and predicting transport and mixing in such flows. 

We are pleased to acknowledge helpful discussions with J.-L. Thiffeault.

\bibliography{reflist}

\appendix

\section{Invariance of Topological Braiding Entropy with Streamline Density}

\begin{figure}
\begin{centering}
\begin{tabular}{c c}
\includegraphics[width=0.36\columnwidth]{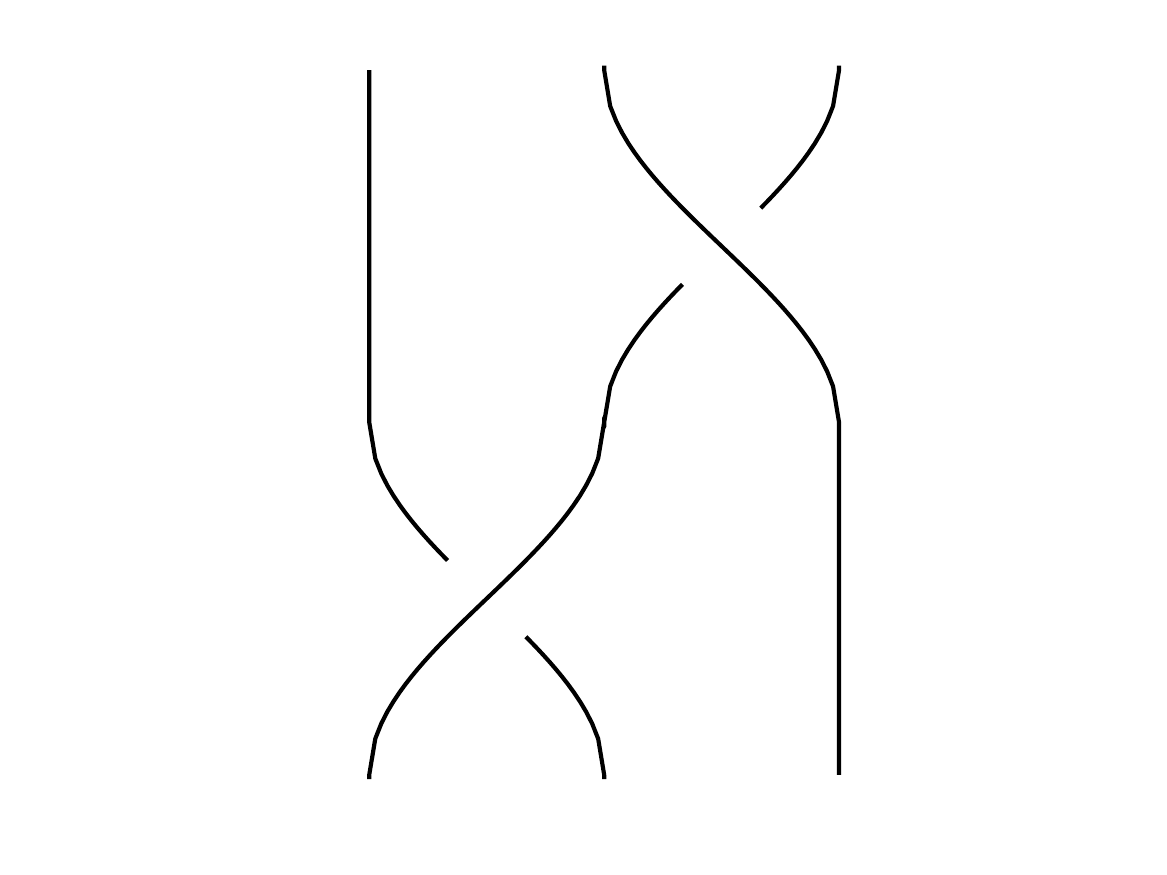}&
\includegraphics[width=0.35\columnwidth]{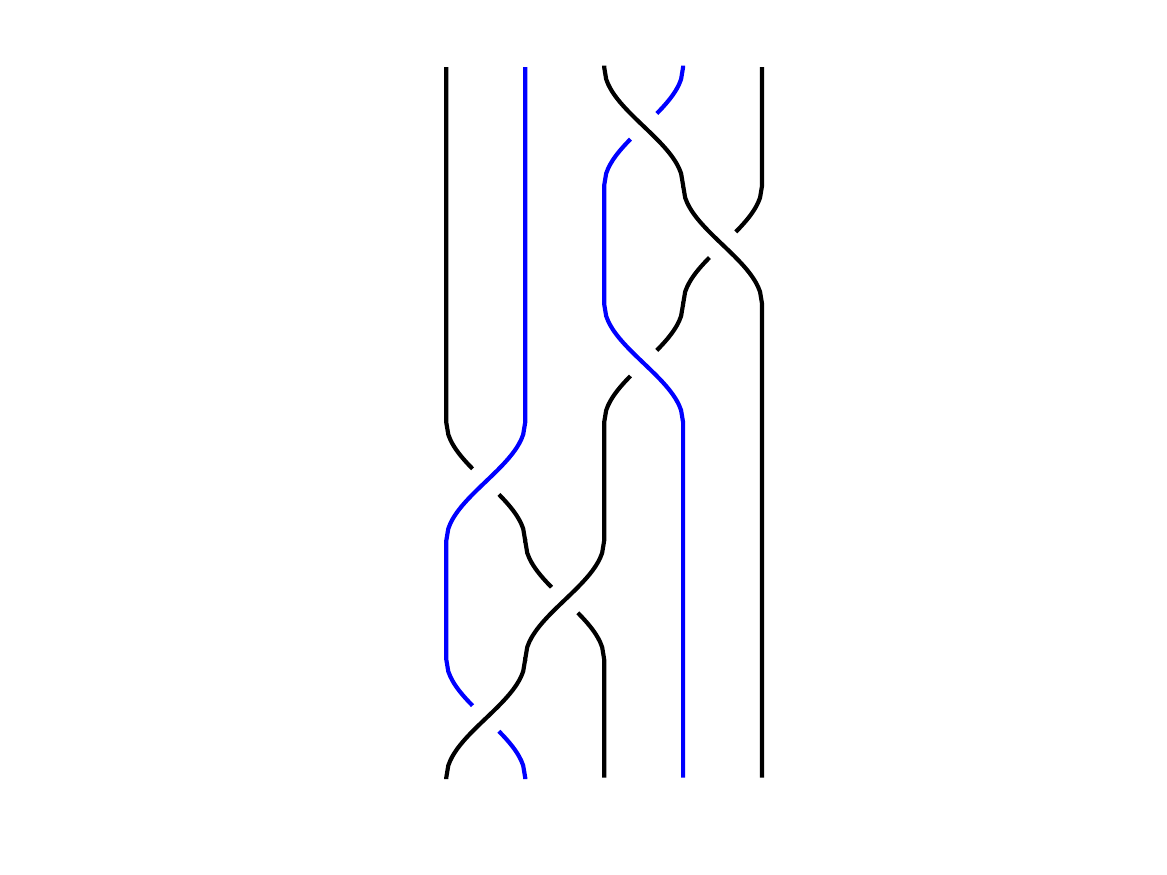}\\
(a) & (b)
\end{tabular}
\end{centering}
\caption{(a) Braid diagram for pigtail braid involving $N_p=3$ streamlines. (b) Braid diagram for pigtail braid involving $N_p=5$ streamlines. Additional streamlines (blue) only undergo trivial braiding and so do not contribute to topological entropy $h$.}
\label{fig:pigtail}
\end{figure}

To show that decreasing streamline spacing $\Delta x_2$ below $\Delta x_2=\ell$ does not impact the braiding exponent in the 1D streamline model, we consider doubling the streamline resolution to $\Delta x_2=\ell/2$, such that the number of streamlines increases from $N_p$ to $2N_p-1$ and the corresponding streamline numbers map as $i\mapsto 2i-1$. This change then corresponds to having 3 neighbouring streamlines rotate together during a clockwise or anti-clockwise reorientation event. Under this change, the clockwise and anticlockwise braid generators are then mapped from the $\Delta x_2=\ell$ streamline model to the $\Delta x_2=\ell/2$ model as
\begin{align}
    \sigma_i^{\pm 1}\mapsto \sigma_{2i-1}^{\pm 1}\sigma_{2i}^{\pm 1}\sigma_{2i-1}^{\pm 1}.\label{eqn:braidmap}
\end{align}
For example, the pigtail braid with $N_p=3$ particles shown in Fig.~\ref{fig:pigtail}a has braid word $\mathbf{b}=\sigma_1\sigma_2^{-1}$, which results in topological entropy $h=\ln[(\sqrt{5}+3)/2]$~\cite{braidbook}. Doubling of the spatial resolution of the streamlines to $N_p=5$ particles via (\ref{eqn:braidmap}) results in the braid diagram shown in shown in Fig.~\ref{fig:pigtail}b with corresponding braid word $\mathbf{b}=\sigma_1\sigma_2\sigma_1\sigma_3^{-1}\sigma_4^{-1}\sigma_3^{-1}$, which has exactly the same topological entropy. This is because the additional streamlines do no contribute to the topological complexity of the braiding motions. As they are added on a lengthscale $\Delta x_2$ less than the correlation length $\ell$, these additional streamlines do not resolve any hitherto hidden stirring actions, they simply move in accordance with the stirring motions resolved by the original set of streamlines. Similar invariance of the topological braiding entropy to streamline resolution resolution below $\Delta x_2=\ell$ is obtained for numerical tests (not shown) of random braiding protocols subject to (\ref{eqn:braidmap}).

\section{Extension of 1D Streamline Model to 2D}

To extend the 1D streamline model introduced in the main body of the paper to 2D, we consider a square array of $N_p=N_x\times N_x$ streamlines in the $x_2-x_3$ plane with spacing $\Delta x_2=\Delta x_3 = \ell$ as shown in Fig.~\ref{fig:2Dbraid}a. Similar to the 1D model, clockwise and anti-clockwise rotations of pairs of adjacent streamlines are chosen at random, but with the variation that adjacent streamlines may now be chosen in both the $x_2$ and $x_3$ directions. Under this 2D model the spatial variance in the $x_2$ and $x_3$ directions then evolve as
\begin{align}
    \sigma^2_{x_2}(t)=\sigma^2_{x_3}(t)=\sigma^2_0+\frac{2}{d}\ell^2\frac{N_b}{N_p}=\sigma^2_0+\frac{2}{d}\ell^2\frac{1}{N_p}\frac{\langle v_1\rangle}{\Delta_L}t,
\end{align}
where $\Delta_L=x_1/N_b$ is the distance between braiding events in the $x_1$ direction. The factor of $2/d$ (where $d=2$ is the Euclidean dimension of the streamline array) associated with the $\ell^2 N_b/N_p$ term arises as each braid event moves two of the $N_p$ streamlines by $\pm\ell$ in the $x_2$ direction $1/d$ of the time and the $x_3$ direction in the other $1/d$. Similarly, the streamline variance due to transverse dispersivity $D_T$ grows as  
\begin{equation}
    \sigma^2_{x_2}(t)=\sigma^2_{x_3}(t)=\sigma^2_0+2D_T t,
\end{equation}
hence the transverse dispersivity is related to the model parameters as
\begin{equation}
    D_T=\frac{\ell^2 \langle v_1\rangle}{d N_p\Delta_L}.\label{eqn:disp}
\end{equation}

\begin{figure}[h]
\begin{centering}
\begin{tabular}{c c}
\includegraphics[width=0.45\columnwidth]{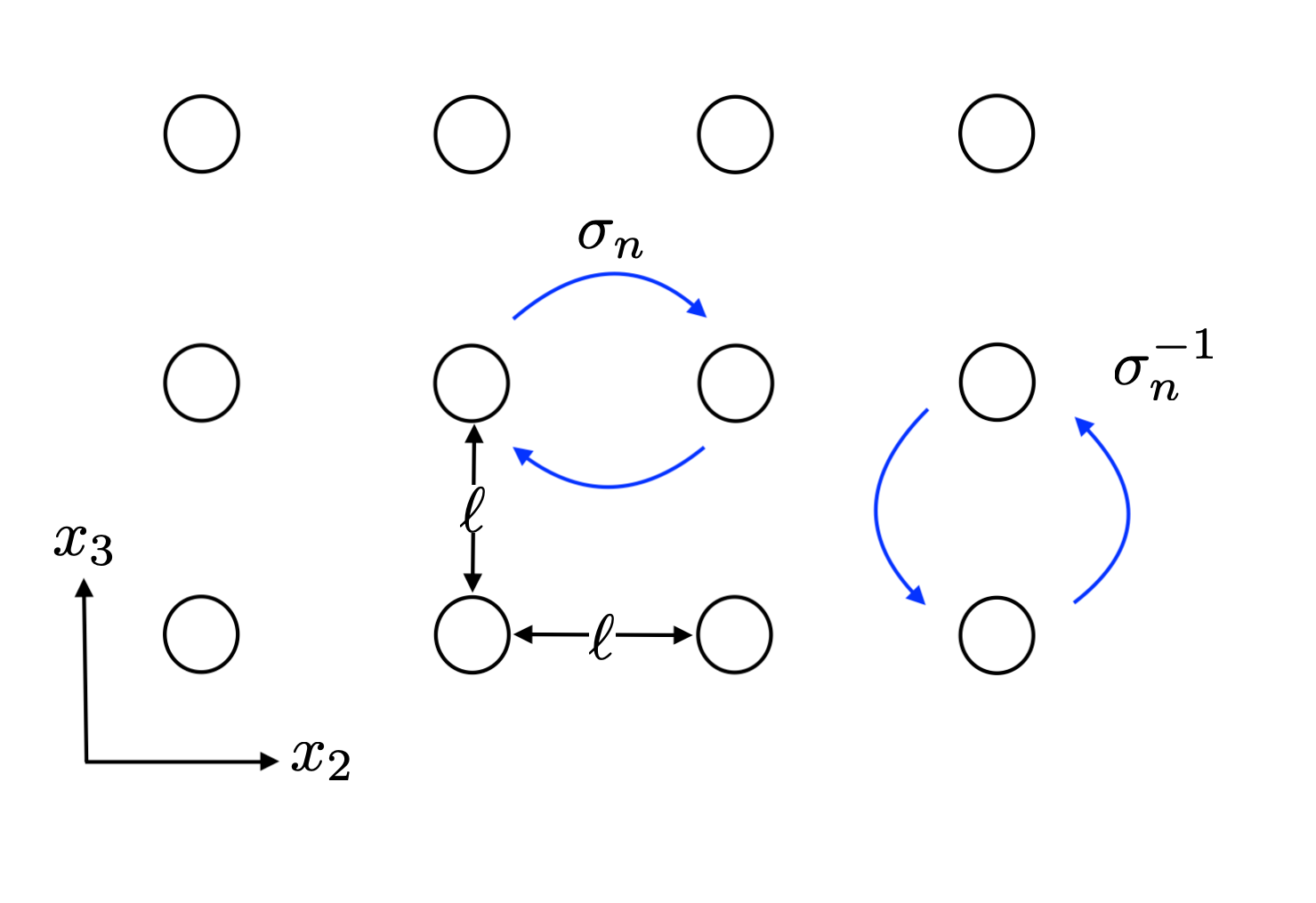}&
\includegraphics[width=0.30\columnwidth]{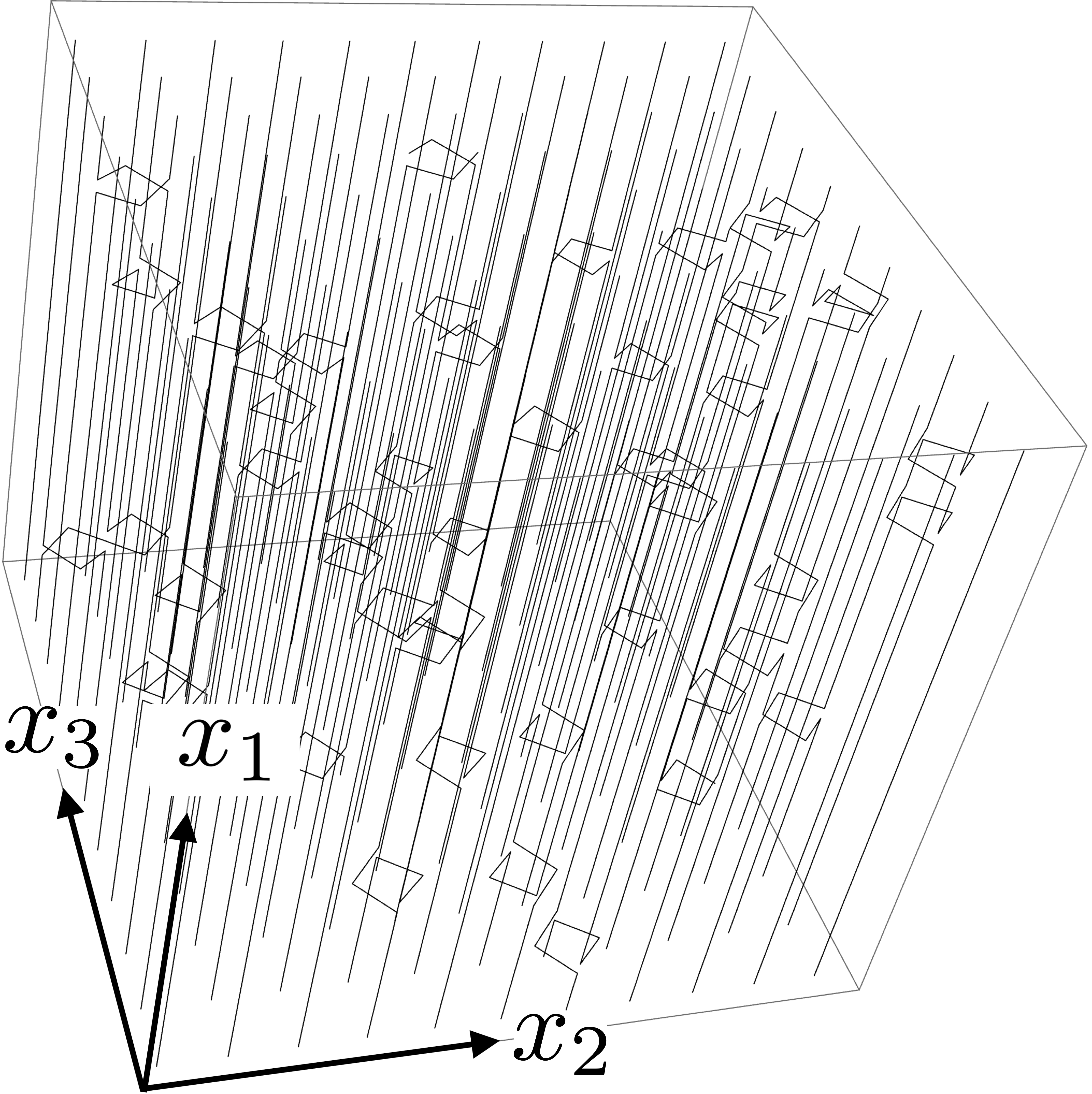}\\
(a) & (b)
\end{tabular}
\end{centering}
\caption{(a) Schematic of streamline braiding in 2D streamline model. (b) Corresponding streamlines for $N_p=N_x\times N_x=10\times 10$ 2D array of streamlines. Helical streamline structures correspond to clockwise and counter-clockwise braiding motions depicted in (a).}
\label{fig:2Dbraid}
\end{figure}

For the 2D streamline model, determination of the sequence of braid generators (braid word) from the sequence of braiding events is more complex. This is because a single braid event involving the exchange of position of two streamlines involves the crossing of these streamlines over many of the other streamlines in the array (depending upon the projection angle used to characterise the crossings~\cite{Thiffeault:2005aa}). To circumvent this problem, for a given sequence of randomly chosen braid events we construct $N_p$ streamline trajectories which resolve the clockwise and anti-clockwise motions of each braiding event. The resultant braid word is computed from the trajectory data using the \emph{databraid} routine in \emph{braidlab}. As the corresponding braid is not closed, the finite time braiding exponent (FTBE)~\cite{Budisic:2015aa}, an analogue of the finite time Lyapunov exponent (FTLE), is used to quantify the degree of entanglement of streamlines. \citet{Budisic:2015aa} show that as the number of braiding events $N_b$ grows, the FTBE converges to the topological entropy of the braid divided by the elapsed time over which $N_b$ has occurred.\\

\begin{figure}
\begin{centering}
\begin{tabular}{c c}
\includegraphics[width=0.36\columnwidth]{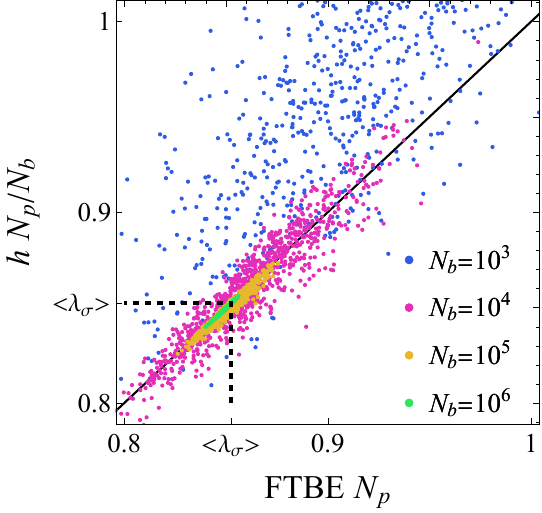}&
\includegraphics[width=0.375\columnwidth]{./ent_converge_2D.pdf}\\
(a) & (b)
\end{tabular}
\end{centering}
\caption{(a) Convergence of scaled FTBE ($N_p$ FTBE) and scaled topological entropy $(h N_p/N_b)$ to the random braid entropy $\langle\lambda_\sigma\rangle$ for the 1D streamline model with the number $N_b$ of braid events for fixed streamline number $N_p=10^3$. Similar results are obtained for larger values of $N_p$. (b) Convergence of scaled FTBE ($\Delta_L/\langle v_1\rangle$ FTBE $N_x$) with braid number $N_b$ for the 2D streamline model for different array dimensions $N_x$ (green $N_x=10$, light blue $N_x=20$, purple $N_x=50$, pink $N_x=100$, red $N_x=200$). Inset: convergence of scaled FTBE for $N_b=10^7$ with streamline array dimension $N_x$ to $\langle\lambda_\sigma\rangle\approx 0.8525$.}
\label{fig:FTBE}
\end{figure}

Fig.~\ref{fig:FTBE}a shows convergence of the scaled FTBE ($\Delta_L/\langle v_1\rangle$ FTBE $N_p$) to the scaled topological entropy ($h N_p/N_b$) with increasing number of braid events $N_b$ for the 1D streamline model over $10^3$ realisations of random braiding sequences. As the FTBE is normalised by the advection time $t$ over which the braiding occurs, the FTBE has units of inverse time and so the prefactor $\Delta_L/\langle v_1\rangle$ renders the scaled FTBE dimensionless. For each braid word, the corresponding streamlines are constructed (see Fig.~\ref{fig:2Dbraid}b) that resolve the clockwise and anti-clockwise rotations, and the \emph{databraid} routine is used to compute the FTBE from these streamlines. The topological entropy is also computed in the usual manner via the \emph{annbraid} routine in \emph{braidlab}. In the limit of large particle numbers $N_p$ and braid events $N_b$, the scaled FTBE and topological entropy converge to each other and the random braiding exponent $\langle\lambda_\sigma\rangle$.\\

For the 2D streamline model, we construct the streamline trajectories shown in Fig.~\ref{fig:2Dbraid}b from the sequence of braid events as described above and compute the FTBE using the \emph{databraid} routine. Fig.~\ref{fig:FTBE}b shows convergence of the scaled FTBE ($\Delta_L/\langle v_1\rangle$ FTBE $N_x$) to the random braid entropy $\langle\lambda_\sigma\rangle$ with increasing number of braid events $N_b$ and particles $N_p=N_x\times N_x$ for the 2D streamline model over 10 realisations of random braiding sequences. Hence for the scaled FTBE is related to the random braid entropy as
\begin{equation}
\lim_{N_b\rightarrow\infty}\lim_{N_p\rightarrow\infty}\frac{\Delta_L}{\langle v_1\rangle}\,\text{FTBE}\,N_x=\langle\lambda_\sigma\rangle.
\end{equation}
\citet{Budisic:2015aa} show that in limit of long times and large particle numbers the FTBE converges to the Lyapunov exponent $\hat{\lambda}_\infty$, hence the dimensionless Lyapunov exponent $\lambda_\infty\equiv\hat{\lambda}_\infty \ell/\langle v_1\rangle$ converges to the scaled FTBE as
\begin{equation}
    \frac{\ell}{\langle v_1\rangle}\lim_{N_b\rightarrow\infty}\lim_{N_p\rightarrow\infty}\,\text{FTBE}=\lambda_\infty.
\end{equation}
Squaring this expression yields
\begin{equation}
    \lambda_\infty^2=\frac{\ell^2}{\langle v_1\rangle^2}\,\text{FTBE}^2=\langle\lambda_\sigma\rangle^2\frac{\ell^2}{\Delta_L^2}\frac{1}{N_p}
\end{equation}
and substitution of (\ref{eqn:disp}) yields
\begin{equation}
    \lambda_\infty^d=\langle\lambda_\sigma\rangle^d\frac{d D_T}{\langle v_1\rangle\Delta_L}=\langle\lambda_\sigma\rangle^d\left(\frac{\ell}{\Delta_L}\right)^{d-1}\frac{d}{Pe_T},\quad d=1,2.
\end{equation}
Hence the 2D streamline model recovers equation (7) in the main body of the paper relating the Lyapunov exponent to the transverse dispersivity.

\section{Extension of 2D Streamline Model to 2D Random Walk Model}

To extend the 2D streamline model to a 2D random walk model, following \cite{braidbook} we generate $N_p$ random streamlines in the domain $\Omega:(x_1,x_2,x_3)\in[0,\infty)\times[0,L]\times[0,L]$. As shown in Fig.~\ref{fig:2DRandom}a), the streamlines are all initiated at random locations within the $x_1=0$ plane in $\Omega$ and propagate longitudinally in the $x_1$ direction by steps of size $\Delta_L$. During each step, the streamlines also make a step of magnitude $\Delta_T$ at a random direction (chosen such that the streamline remains confined within $\Omega$) in the $x_2-x_3$ plane.

The major differences between this random streamline model and the 1D and 2D streamline models discussed in the previous section is that (i) streamlines make a random walk rather than being confined to braiding within the streamline array, and (ii) all the streamlines undergo simultaneous braiding events similar to real flows, rather than the single braiding events encoded in the models derived from braid generators. We show below that these changes make no material differences to the transverse dispersion/fluid stretching relationship (9) derived in the main body of this paper.

Although confinement of streamlines to the domain $\Omega$ leads to zero transverse dispersivity, Fig.~\ref{fig:2Dbraid}b shows that the braiding entropy of these random streamlines approaches that of unconfined streamlines as $N_p$ becomes large. Hence the braiding entropy of this confined streamline model may be compared to the dispersivity of the unconfined streamline model. For the unconfined model, the transverse variances grow as
\begin{equation}
    \sigma^2_{x_2}(t)=\sigma^2_{x_3}(t)=\sigma^2_0+\frac{\Delta_T^2}{2}\frac{x_1}{\Delta_L}=\sigma^2_0+\frac{\Delta_T^2}{2\Delta_L} \langle v_1\rangle t,
\end{equation}
as the number of random steps of size $\Delta_T$ is $x_1/\Delta_L=\langle v_1\rangle t/\Delta_L$. As these variances both grow linearly as $2D_T t$, the relationship between transverse dispersivity $D_T$ and the random walk parameters is then
\begin{equation}
    D_T=\frac{\Delta_T^2}{4\Delta_L}\langle v_1\rangle.\label{eqn:disp2}
\end{equation}

\begin{figure}
\begin{centering}
\begin{tabular}{c c c}
\includegraphics[width=0.22\columnwidth]{./2D_random_walk.png}&
\includegraphics[width=0.36\columnwidth]{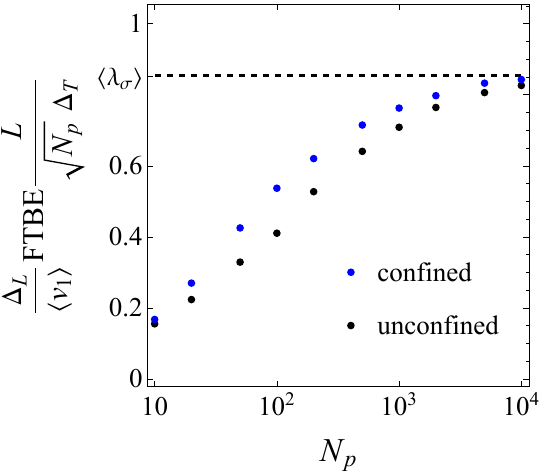}&
\includegraphics[width=0.36\columnwidth]{./converge_2D_deltaT.pdf}\\
(a) & (b) & (c)
\end{tabular}
\end{centering}
\caption{(a) Schematic of 2D random walk streamline model. (b) Convergence of scaled FTBE (\ref{eqn:FTBEscaled}) for the unconfined (black dots) and confined (blue dots) 2D random walk models to the random braid entropy $\langle\lambda_\sigma\rangle$ with the number $N_p$ of of streamlines and $N_{x_1}=10^3$ random steps. (c) Convergence of scaled FTBE (\ref{eqn:FTBEscaled}) to the random braid entropy $\langle\lambda_\sigma\rangle$ with the number $N_p$ of of streamlines for different values of $\Delta_T/L$.}
\label{fig:2DRandom}
\end{figure}

Fig.~\ref{fig:2DRandom} shows that as the number of streamlines $N_p$ and longitudinal increments $N_{x_1}=x_1/\Delta_L$ becomes large, the scaled FTBE converges to the random braid entropy as
\begin{equation}
\lim_{N_p\rightarrow\infty}\lim_{N_{x_1}\rightarrow\infty}\frac{\Delta_L}{\langle v_1\rangle}\text{FTBE}\frac{1}{\sqrt{N_p}}\frac{L}{\Delta_T}=\langle\lambda_\sigma\rangle,\label{eqn:FTBEscaled}
\end{equation}
suggesting that braiding in the random streamline model is also governed by this universal constant. The different scaling of the FTBE for this model arises because all $N_p$ streamlines braid at each longitudinal increment $\Delta_L$, whereas for the 1D and 2D streamlines, only one braiding event occurs across all $N_p$ streamlines respectively. As the streamlines in this random walk model all undergo independent random walks, then in the same manner as the 1D and 2D streamline models, the average spacing between nearest streamlines (given by $L/\sqrt{N_p}$) is equivalent to the correlation length scale $\ell$ of the flow. Squaring (\ref{eqn:FTBEscaled}) and substitution of (\ref{eqn:disp2}) yields 
\begin{equation}
\frac{\Delta_L^2}{\langle v_1\rangle^2}\text{FTBE}^2=\langle\lambda_\sigma\rangle^2 N_p\frac{\Delta_T^2}{L^2}=\langle\lambda_\sigma\rangle^2 \frac{N_p}{L^2}\frac{d^2 D_T\Delta_L}{\langle v_1\rangle^2},
\end{equation}
Hence the relationship between Lyapunov exponent and transverse dispersivity is then
\begin{equation}
    \lambda_\infty^2=\langle \lambda_\sigma\rangle^2\frac{d^2 D_T}{\langle v_1\rangle\Delta_L}=\langle \lambda_\sigma\rangle^2\frac{\ell}{\Delta_L}\frac{d^2}{Pe_T}.
\end{equation}

\section{Random Conductivity Fields}

The random fields $f(\mathbf{x})=\ln K(\mathbf{x})$ used for the hydraulic conductivity fields $k_0(\mathbf{x})$, $k_\delta(\mathbf{x})$ are given by the Fourier modes
\begin{equation}
\begin{split}
f(\mathbf{x})&=\Sigma^2_{\ln k}\sum_{n=1}^{N_i}\sum_{i,j,k}^{N} \frac{A_{n,ijk}}{\sqrt{i^2+j^2+k^2}}
\cos(2\pi i (x_1+\phi^1_{n,ijk}))\\
&\cos(2\pi j (x_2+\phi^2_{n,ijk}))\cos(2\pi k (x_3+\phi^3_{n,ijk})),
\end{split}
\end{equation}
where $N=2$ is the number of Fourier modes and so the correlation length of these random fields is $\ell=1/(2N)$. $N_i$ is the number of realisations in each mode, and the coefficients $A_{n,ijk}$ and the phase angles $\phi_{n,ijk}$ are uniformly distributed random variables in $[0,1]$ and the coefficient $\Sigma^2_{\ln k}$ is chosen such that the log-variance $||f^2||_\mathcal{D}=\sigma^2_{\ln k}=4$. We note that $N=2$ is required to generate chaotic dynamics as symmetric conductivity fields (corresponding to $N=1$) do not generate chaotic dynamics due to the symmetry of the resultant velocity field.

\section{Darcy Flow Numerical Solvers and Streamline Tracking}

To solve the divergence-free condition $\nabla\cdot\mathbf{v}=0$ for the Darcy flow given by equation (9) in the main body of the paper over the triply-periodic domain $\mathcal{D}$, we decompose the potential field into mean and fluctuating components respectively as 
\begin{equation}
\phi(\mathbf{x})=\bar\phi(\mathbf{x})+\tilde\phi(\mathbf{x}),\label{eqn:phidecomp}
\end{equation}
and set the mean potential as $\bar\phi=-x_1$. The resulting governing equation for the potential fluctuation $\tilde{\phi}(\mathbf{x})$ is then
\begin{equation}
\nabla\cdot(\mathbf{K}\cdot\nabla\tilde{\phi}(\mathbf{x}))=
\nabla\cdot(\mathbf{K}\cdot\hat{\mathbf{e}}_1).
\end{equation}
This equation is solved to precision $10^{-16}$ on a uniform structured $256^3$ grid using a high resolution eighth-order compact finite difference scheme~\cite{LeLe:1992aa} via an iterative Krylov method. To generate high resolution results and preserve the Lagrangian kinematics of the flow, we use a similar numerical approach to that used in \cite{Lester:2019ab}. Specifically, we perform a triply-periodic 5-th order spline interpolation of the primitive variables $\tilde{\phi}(\mathbf{x})$, $\mathbf{K}(\mathbf{x})$ from their grid values and reconstruct a smooth, continuous potential field $\phi(\mathbf{x})$ according to (\ref{eqn:phidecomp}). The velocity field is then computed analytically from these fields as $\mathbf{v}(\mathbf{x})=-\mathbf{K}(\mathbf{x})\cdot\nabla\phi(\mathbf{x})$, ensuring that the velocity field is triply-periodic and $C_4$ continuous and the velocity gradient is accurately resolved. For the $256^3$ mesh the local relative divergence error 
\begin{equation}
    d(\mathbf{x})=\frac{|\nabla\cdot\mathbf{v}(\mathbf{x})|}{||\nabla\mathbf{v}(\mathbf{x})||},
\end{equation}
of the interpolated velocity field $\mathbf{v}(\mathbf{x})$ is order $10^{-4}$ and the velocity gradient is accurate to order $10^{-3}$. Streamline tracking is then computed by solving the advection equation from the initial Lagrangian coordinate $\mathbf{X}$ as
\begin{equation}
    \frac{d\mathbf{x}}{dt}=\mathbf{v}(\mathbf{x}(t;\mathbf{X})),\quad\mathbf{x}(0;\mathbf{X})=\mathbf{X},\label{eqn:advection}
\end{equation}
via a 5-th order Cash-Karp Runge-Kutta scheme to precision $10^{-14}$. The periodic boundaries allow advection of fluid streamlines over many multiples of the solution domain $\mathcal{D}$, facilitating study of the Lagrangian kinematics over arbitrary distances. Although the corresponding velocity field is periodic in space, when the flow is chaotic the streamlines are aperiodic and eventually sample all of the conductivity field in an ergodic manner.\\

While accurate, this streamline tracking method (along with all numerical schemes which do not explicitly enforce kinematic constraints) has been shown~\cite{Lester:2023aa} to introduce spurious transverse dispersion for the helicity-free flow $h=0$ due to numerical streamlines not follwing their analytic counterparts. To circumvent this problem for the helicity-free case $\delta=0$, we instead solve the invariant streamfunctions $\psi_1(\mathbf{x})$, $\psi_2(\mathbf{x})$ for the velocity field $\mathbf{v}(\mathbf{x})=\nabla\psi_1(\mathbf{x})\times\nabla\psi_2(\mathbf{x})$ via the following governing equations~\cite{Lester:2022aa} to precision $10^{-16}$ using the same finite-difference method as described above:
\begin{align}
\nabla^2\psi_1(\mathbf{x})-\nabla f(\mathbf{x})\cdot\nabla\psi_1(\mathbf{x})=S_1(\psi_1,\psi_2),\\
\nabla^2\psi_2(\mathbf{x})-\nabla f(\mathbf{x})\cdot\nabla\psi_2(\mathbf{x})=S_2(\psi_1,\psi_2),
\label{eqn:sfuncs}
\end{align}
where $f=\ln k$ and 
\begin{align}
S_1=\frac{(\mathbf{B}\times\psi_2)\cdot(\nabla\psi_1\times\nabla\psi_2)}{|\nabla\psi_1\times\nabla\psi_2|} &&
S_2=\frac{(\mathbf{B}\times\psi_1)\cdot(\nabla\psi_1\times\nabla\psi_2)}{|\nabla\psi_1\times\nabla\psi_2|},
\end{align}
and
\begin{equation}
    \mathbf{B}\equiv(\nabla\psi_1\cdot\nabla)\nabla\psi_2)-(\nabla\psi_2\cdot\nabla)\nabla\psi_1).
\end{equation}
Similar to the Darcy equation, continuous streamfunctions $\psi_1(\mathbf{x})$, $\psi_2(\mathbf{x})$ are reconstructed from grid data using triply-periodic splines and the velocity field is computed analytically from these streamfunctions. As shown in \cite{Lester:2022aa}, this method yields the same velocity field (to within numerical error) as that given by direct solution of the Darcy equation.\\

Each family of streamfunctions $\psi_i$ is comprised of a \emph{foliation} of non-intersecting streamsurfaces $\psi_i=\text{const.}$ that span the flow domain and constrain the Lagrangian kinematics of the flow. This flow structure is non-chaotic as the advection equation (\ref{eqn:advection}) simplifies to
\begin{equation}
\frac{ds}{dt}=v(s;\psi_1(\mathbf{X}),\psi_2(\mathbf{X}),\quad s(t=0;\mathbf{X})=0,
\label{eqn:advect}
\end{equation}
where $s$ is the distance travelled along a streamline of a tracer particle with initial position $\mathbf{X}$. The velocity magnitude $v(s;\psi_1(\mathbf{X}),\psi_2(\mathbf{X})$ at the intersection of the streamsurfaces $\psi_1(\mathbf{x})=\psi_1(\mathbf{X})$, $\psi_2(\mathbf{x})=\psi_2(\mathbf{X})$ only varies with $s$. Equation (\ref{eqn:advect}) is \emph{integrable} in that $\psi_1$, $\psi_2$ represent two invariants of the flow in the 3D domain, resulting in only one degree of freedom (distance) for streamlines of the flow to explore. \citet{Lester:2023aa} subsequently proved that the confinement of streamlines to streamsurfaces $\psi_1$, $\psi_2$ prohibits transverse hydrodynamic dispersion in the asymptotic limit if the isotropic conductivity field $k(\mathbf{x})$ is smooth, regardless of the heterogeneity or conductivity structure of the medium at the Darcy scale. For helicity-free flow, we perform streamline tracking via numerical integration of (\ref{eqn:advect}) to precision $10^{-8}$. This approach ensures numerical streamlines follow their analytic counterparts and so enforces zero transverse dispersion and prevents the non-trivial braiding of streamlines that lead to chaotic advection.

\section{Calculation of Transverse Dispersivity}

\begin{figure}
\begin{centering}
\begin{tabular}{c c}
\includegraphics[width=0.35\columnwidth]{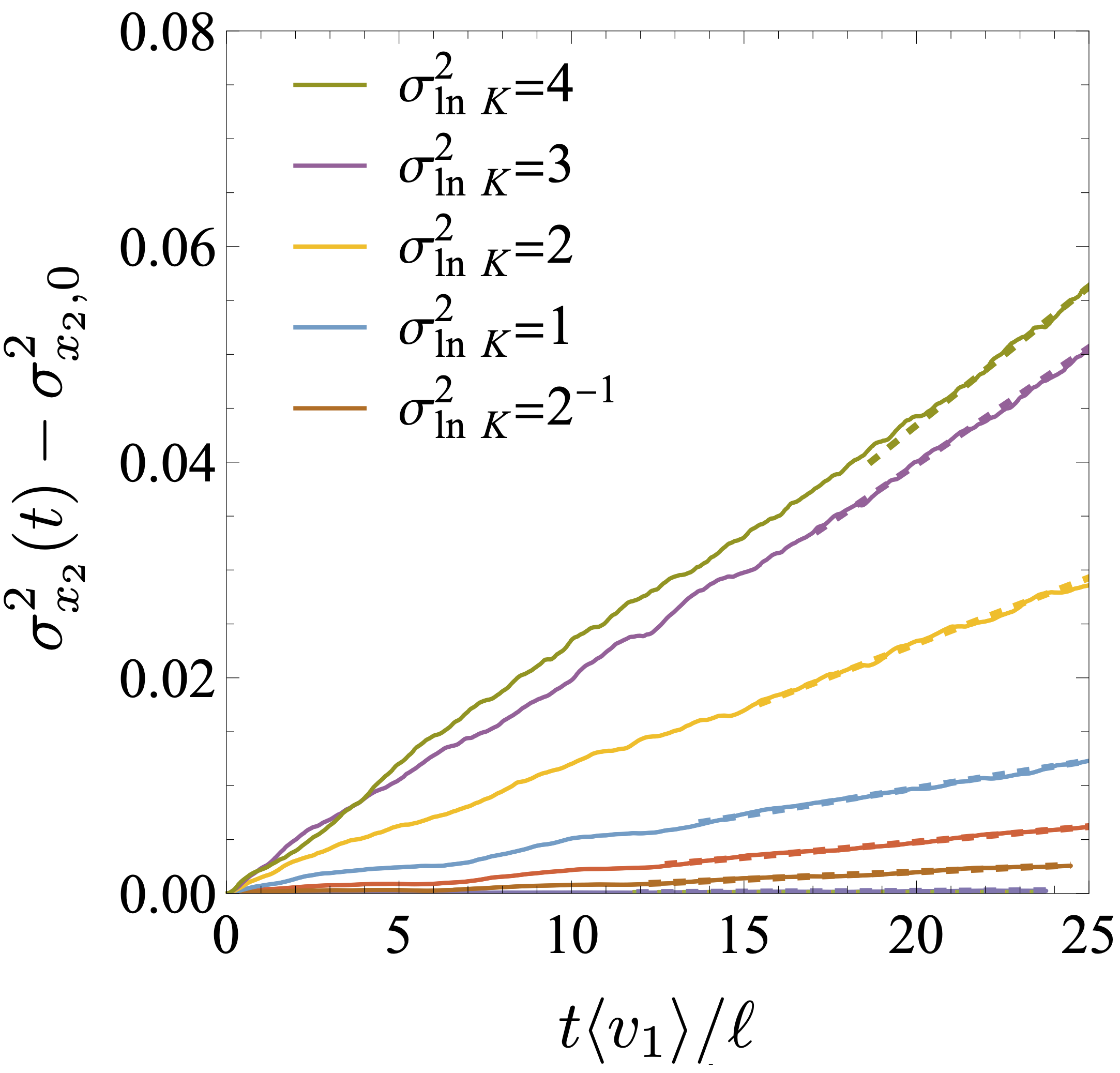}&
\includegraphics[width=0.35\columnwidth]{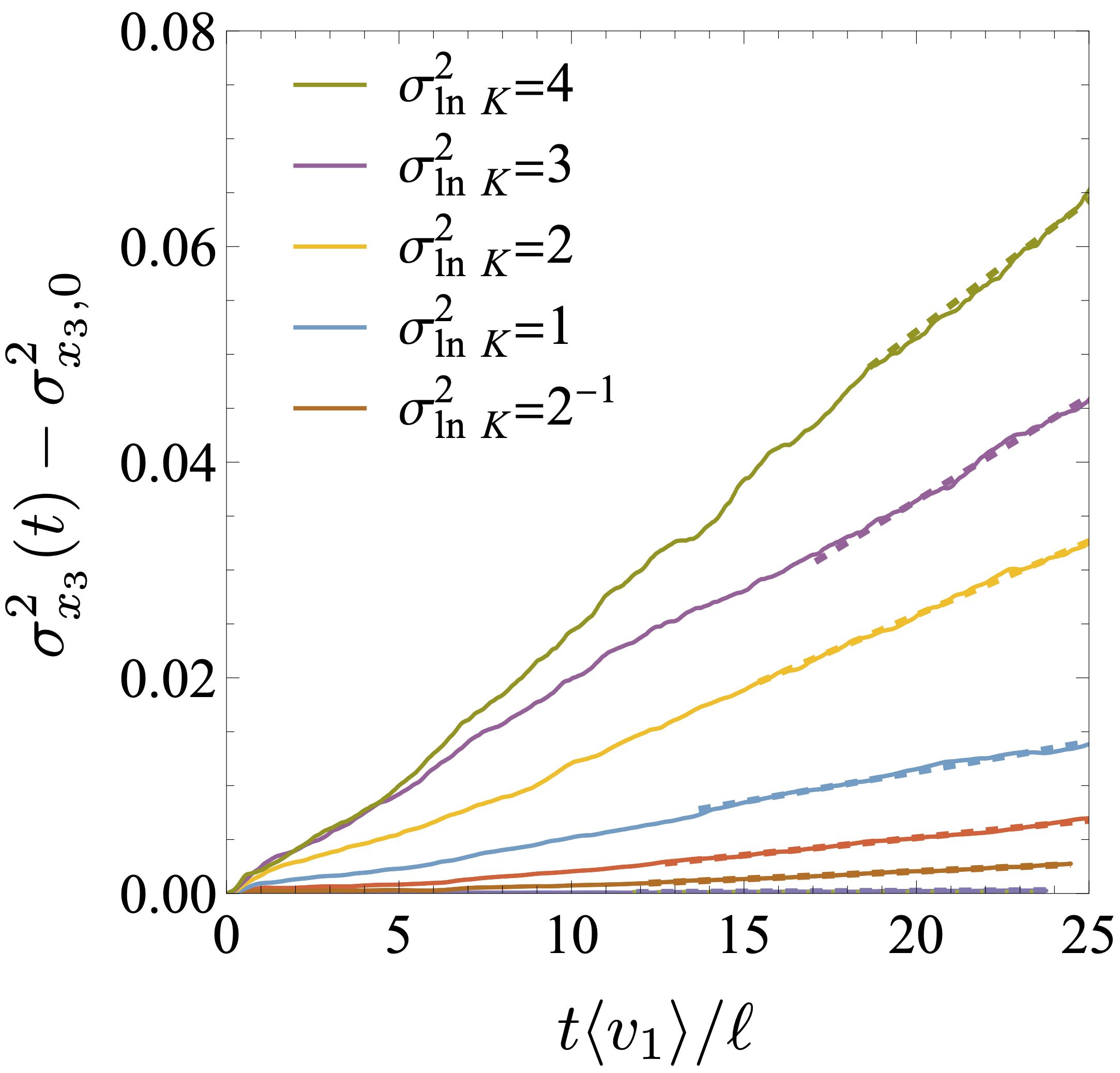}\\
(a) & (b)\\
\includegraphics[width=0.38\columnwidth]{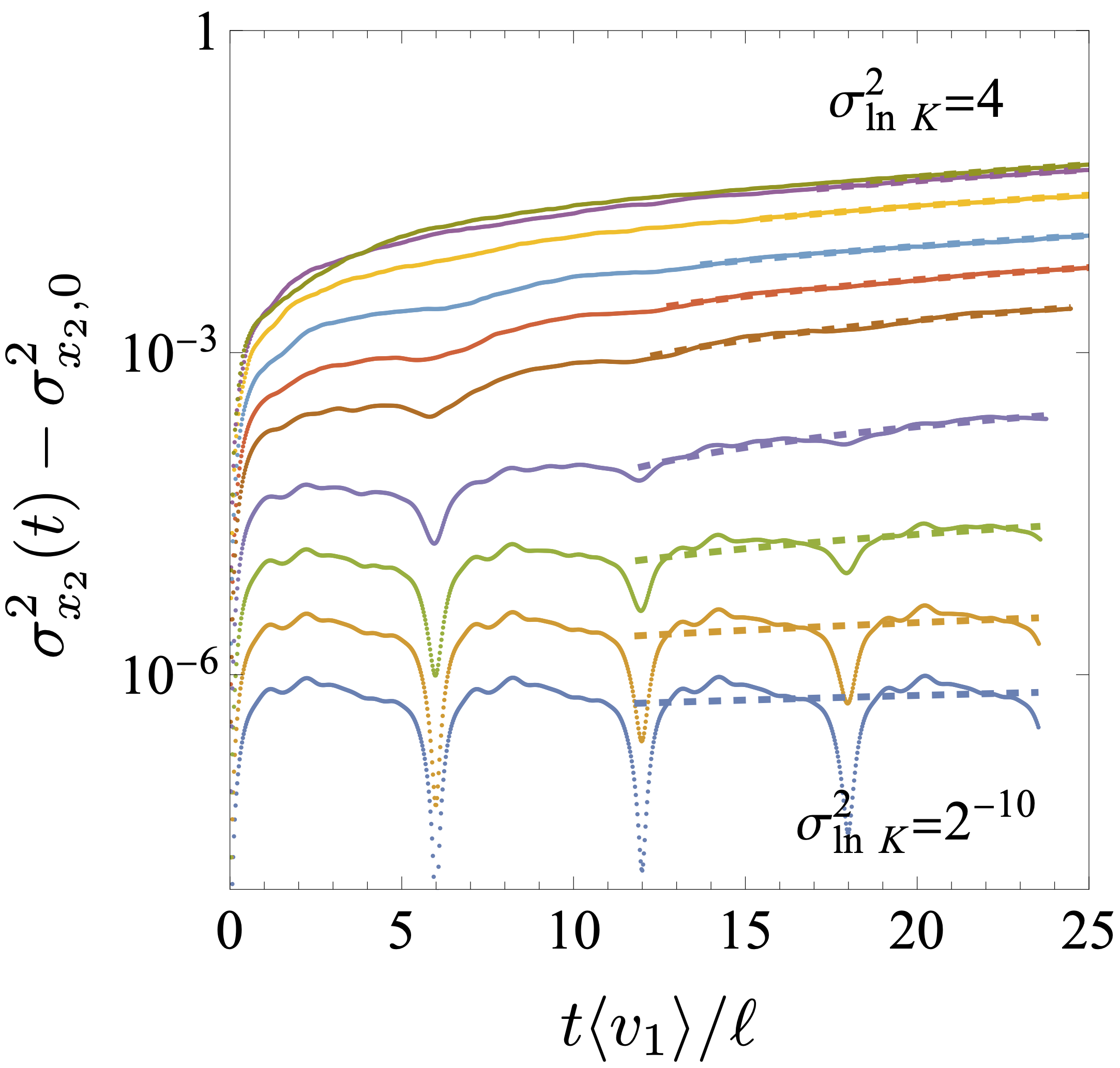}&
\includegraphics[width=0.38\columnwidth]{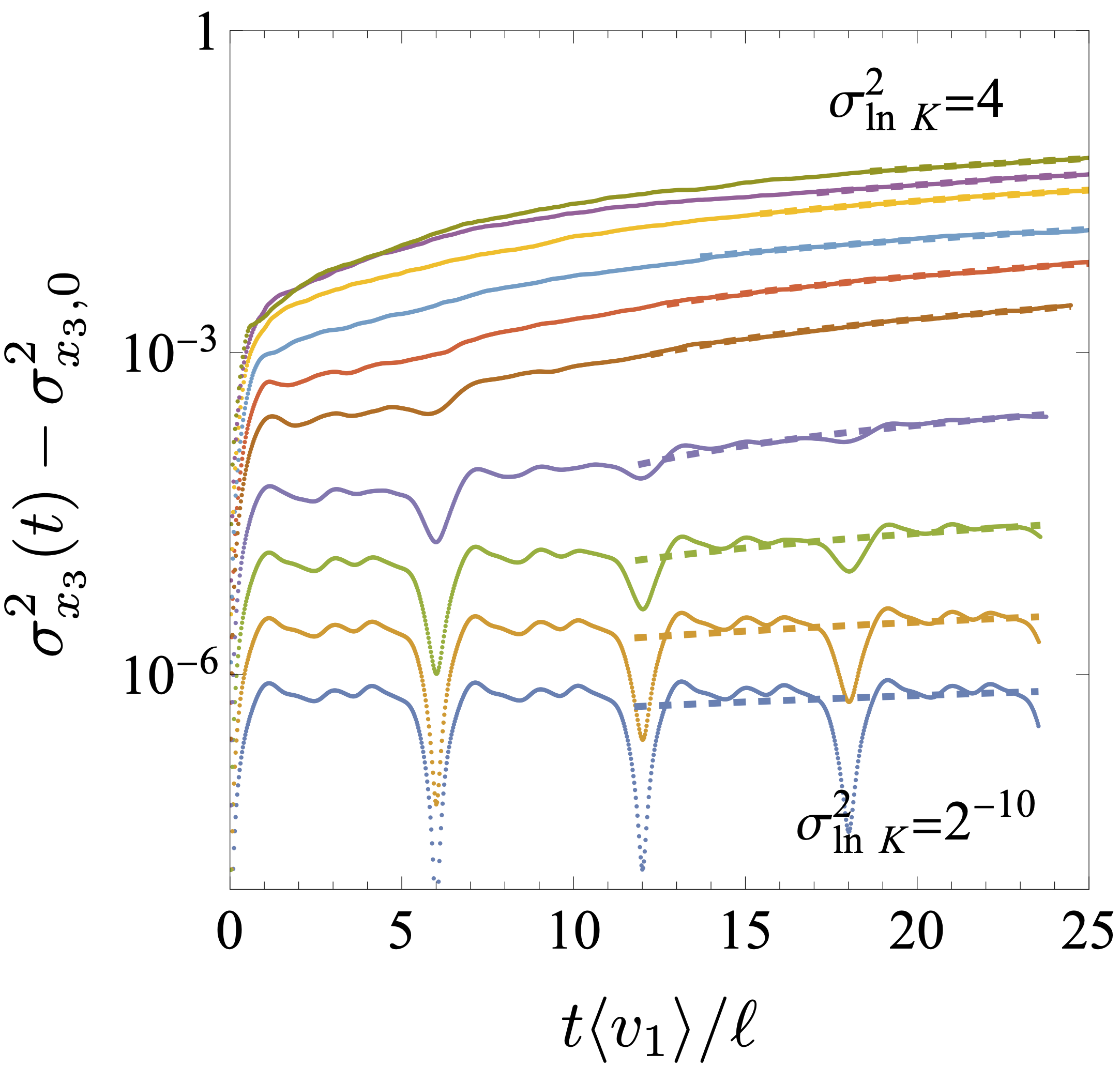}\\
(c) & (d)
\end{tabular}
\end{centering}
\caption{Evolution of transverse scalar variances $\sigma_{x_2}^2(t)$ (left), $\sigma_{x_3}^2(t)$ (right) with dimensionless travel time $t\langle v\rangle/\ell$ in linear (top) and logarithmic (bottom) scales for steady heterogeneous anisotropic 3D Darcy flow for various values of log-conductivity variance $\sigma_{\ln K}^2$. Fitted linear trend (dashed lines) at late times is used to estimate transverse dispersivity.}
\label{fig:variances}
\end{figure}

For both the steady 3D Darcy flow and the unsteady 2D time-periodic random sine flow (TPRSF), the transverse dispersivity is determined by tracking $N_p=10^3$ streamlines or pathlines respectively over $10^3$ traverses of the periodic domain $\mathcal{D}$ seeded from random locations within $\mathcal{D}$ or $10^3$ flow periods $T$. From the 3D streamlines and the 2D pathlines, the transverse variances are computed as
\begin{align}&\sigma_{x_j}^2(t)=\frac{1}{N_p}\sum_{i=1}^{N_p}(x_{j,i}(t)-\langle x_j\rangle(t))^2,\quad \langle x_j\rangle(t)=\frac{1}{N_p}\sum_{i=1}^{N_p}x_{j,i}(t),
\end{align}
where $j=1,2$ for the TPRSF and $j=2,3$ for the 3D Darcy flow. The transverse dispersivities are then related to the asymptotic variance growth as
\begin{align}
    &D_{jj}=\frac{1}{2}\lim_{t\rightarrow\infty}\frac{d\sigma_{x_j}^2}{dt},
\end{align}
To estimate this asymptotic growth for both the TPRSF and the steady 3D Darcy flow we fit a linear function to the asymptotic variance data over two periods of the flow as shown in Fig.~\ref{fig:variances} for the 3D Darcy flow. Note that the variance data in Fig.~\ref{fig:variances} has been truncated to 20 correlation lengths for illustrative purposes. For both the TPRSF and the steady 3D Darcy flow the transverse dispersivity $D_T$ is then computed as the average of the $D_{jj}$ dispersivities.

As shown in Fig.~\ref{fig:variances} for the anisotropic Darcy flow with variable heterogeneity, for small conductivity variance $\sigma^2_{\ln K}\\ 1$, the transverse variances exhibit slow growth and periodic oscillations as the streamlines of the flow are nearly periodic. With increasing heterogeneity $\sigma^2_{\ln K}$, these orbits lose periodicity and ergodically explore the flow domain, resulting in stronger growth of the transverse variances. 

\section{Computation of Lyapunov Exponents}

For the TPRSF, the Lyapunov exponent of these flows is computed by directly integrating the deformation gradient tensor $\mathbf{F}(\mathbf{X},t)$ along pathlines with Lagrangian coordinate $\mathbf{X}$ as
\begin{equation}
\frac{d\mathbf{F}(\mathbf{X},t)}{dt}=\boldsymbol\epsilon(\mathbf{X},t)\cdot\mathbf{F}(\mathbf{X},t),\quad \mathbf{F}(\mathbf{X},0)=\mathbf{1},\label{eqn:deformX},
\end{equation}
where $\boldsymbol\epsilon(\mathbf{X},t)\equiv\nabla\mathbf{v}(\mathbf{x,t})^\top\big|_{\mathbf{x}=\mathbf{x}_0(\mathbf{X},t)}$ and $\mathbf{x}_0(\mathbf{X},t)$ is the position of a streamline initially at position $\mathbf{X}$ at time $t=0$. The simplicity of the TPRSF means that the evolution equation (\ref{eqn:deformX}) can be solved analytically over each flow period $T$, and the finite time Lyapunov exponent for each streamline computed as
\begin{equation}
    \hat{\lambda}(t,\mathbf{X})=\frac{1}{t}\ln\nu(t,\mathbf{X}),
\end{equation}
where $\nu(t,\mathbf{X})$ is the leading eigenvalue of the Cauchy-Green tensor $\mathbf{F}(\mathbf{X},t)^\top\cdot \mathbf{F}(\mathbf{X},t)$. The infinite-time Lyapunov exponent is then computed via the ensemble average
\begin{equation}
\hat{\lambda}_\infty=\lim_{t\rightarrow\infty}\langle\hat{\lambda}(t,\mathbf{X})\rangle.
\end{equation}

The Lyapunov exponent for the steady 3D Darcy flows is computed via the moving and rotating \emph{Protean} coordinate frame as is detailed in \cite{Lester:2018aa} and briefly summarised as follows. The Protean coordinate frame $\mathbf{x}^\prime$ is related to the Eulerian frame $\mathbf{x}$ as $\mathbf{x}^\prime(t)=\mathbf{Q}(t)\cdot(\mathbf{x}-\mathbf{x}_0(t))$, where $\mathbf{x}_0(t)$ is the position of a fluid tracer particle at time $t$ with initial position $\mathbf{X}$, and $\mathbf{Q}(t)$ is a time-dependent orthogonal rotation matrix. The Protean velocity gradient tensor $\boldsymbol\epsilon^\prime(\mathbf{x}^\prime,t)$ is related to the Lagrangian velocity gradient tensor $\boldsymbol\epsilon(\mathbf{X},t)$ as
\begin{equation}
\boldsymbol\epsilon^\prime(\mathbf{X}^\prime,t)=\mathbf{Q}^\top(t)\cdot\boldsymbol\epsilon(\mathbf{X}^\prime,t)\cdot\mathbf{Q}(t)+\dot{\mathbf{Q}}^\top(t)\cdot\mathbf{Q}(t),\label{eqn:Protean_eps}
\end{equation}
where $\boldsymbol\epsilon(\mathbf{X},t)\equiv\nabla\mathbf{v}(\mathbf{x})^\top\big|_{\mathbf{x}=\mathbf{x}_0(\mathbf{X},t)}$. The rotational matrix $\mathbf{Q}(t)$ aligns the $x_1^\prime$ coordinate with the local velocity direction $\mathbf{v}/v$, and for steady flow the Protean coordinate system is a streamline coordinate system. The rotation $\mathbf{Q}(t)$ is comprised of two subrotations as $\mathbf{Q}(t)=\mathbf{Q}_2(t)\cdot\mathbf{Q}_1(t)$, where the first rotation $\mathbf{Q}_1(t)$ aligns $x_1^\prime$ with $\mathbf{v}/v$ and so renders the $\epsilon_{21}^\prime$ and $\epsilon_{31}^\prime$ elements of the Protean velocity gradient tensor $\boldsymbol\epsilon^\prime(t)$ to be zero~\cite{Lester:2018aa}. The second rotation $\mathbf{Q}_2(t)$ about the axis $\mathbf{x}^\prime$ in the streamwise direction is chosen such the remaining lower triangular element $\epsilon_{23}^\prime$ is also zero~\cite{Lester:2018aa}, rendering the Protean velocity gradient tensor upper triangular:
\begin{equation}
    \boldsymbol\epsilon^\prime=\left(
\begin{array}{ccc}
     \epsilon^\prime_{11} & \epsilon^\prime_{12} & \epsilon^\prime_{13} \\
    0 & \epsilon^\prime_{22} & \epsilon^\prime_{23}\\
    0 & 0 & \epsilon^\prime_{33}
\end{array}
    \right).
\end{equation}
This greatly simplifies solution of the deformation gradient tensor $\mathbf{F}^\prime(\mathbf{X}^\prime,t)$ as
\begin{equation}
\frac{d\mathbf{F}^\prime(\mathbf{X}^\prime,t)}{dt}=\boldsymbol\epsilon^\prime(\mathbf{X}^\prime,t)\cdot\mathbf{F}^\prime(\mathbf{X}^\prime,t),\quad \mathbf{F}^\prime(\mathbf{X}^\prime,0)=\mathbf{1},\label{eqn:deform}
\end{equation}
which is related to its Lagrangian counterpart as
\begin{equation}
\mathbf{F}^\prime(\mathbf{X}^\prime,t)=\mathbf{Q}^\top(t)\cdot\mathbf{F}(\mathbf{X}^\prime,t)\cdot\mathbf{Q}(0),
\end{equation}
and $\mathbf{F}^\prime(\mathbf{X}^\prime,t)$ is also upper triangular. The major advantage of the Protean frame is that the diagonal components $\epsilon_{ii}^\prime$ and $F_{ii}^\prime$ represent the principal stretches of material elements, whereas the non-zero off-diagonal components $\epsilon_{ij}^\prime,\,\,j>i,$ represent shear deformations. This means that in the Protean frame, the hyperbolic stretches that directly govern the Lyapunov exponents of the flow are not conflated with other deformations such as shears and rotations, and so may be directly measured from the diagonal components $\epsilon_{ii}^\prime$. Indeed, the diagonal elements of $\mathbf{F}^\prime(\mathbf{X}^\prime,t)$ grow exponentially with the diagonal elements of $\boldsymbol\epsilon^\prime(t)$ as
\begin{equation}
F_{ii}^\prime(t)=\exp\left(\int_0^t dt^\prime \epsilon_{ii}^\prime(t^\prime)\right),\quad i=1:3.\label{eqn:Fii}
\end{equation}
From (\ref{eqn:Fii}), fluid stretching in the streamwise direction simply fluctuates with the local velocity as $F_{11}^\prime(t)=v(t)/v(0)$ but do not grow in random stationary flows as $\langle F_{11}^\prime\rangle\sim 1$. Conversely, the transverse stretches $F_{22}^\prime$ and $F_{33}^\prime$ respectively grow and decay exponentially according to the Lyapunov exponent as $\langle F_{22}^\prime\rangle=1/\langle F_{33}^\prime\rangle\sim\exp(\hat{\lambda}_\infty t)$.\\

Due to linearity of the Darcy equation, we non-dimensionalise the velocity gradient in terms of the mean velocity and correlation length as 
\begin{equation}
\tilde{\boldsymbol\epsilon}^\prime(\mathbf{X}^\prime,t)\equiv\frac{\ell}{\langle v\rangle}\boldsymbol\epsilon^\prime(\mathbf{X}^\prime,t),\label{eqn:nondimeps}
\end{equation}
Using this non-dimensionalisation the dimensionless Lyapunov exponent $\lambda_\infty$ can be quantified directly in terms of the ensemble averages $\langle \tilde{\epsilon}_{ii}^\prime\rangle, i=2,3$ as
\begin{equation}
\lambda_\infty=\langle\tilde{\epsilon}^\prime_{22}\rangle=-\langle\tilde{\epsilon}^\prime_{33}\rangle,\label{eqn:Lyapunov}
\end{equation}
with $\langle\tilde{\epsilon}_{11}(t)\rangle=0$ and the divergence-free condition manifests as $\sum_{i}\tilde{\epsilon}_{ii}^\prime=0$. Hence to leading order chaotic advection in steady 3D flow is characterised by the Lyapunov exponent $\lambda_\infty$.\\

Previous studies~\cite{Le-Borgne:2008aa,Le-Borgne:2008ab} have established that for steady flows in random media at both the pore- and Darcy scales, the velocity magnitude $v$ decorrelates along streamlines in space rather than time (whereas the velocity magnitiude exhibits strong intermittency in time), such that the velocity follows a spatial Markovian process with respect to the correlation length $\ell$. \citet{Lester:2018aa} showed that the velocity gradient $\epsilon$ also follows a spatial Markovian process and also decorrelates with respect to the correlation length $\ell$. The Protean velocity gradient $\boldsymbol\epsilon^\prime$ is computed along the $10^3$ streamlines of the flow for a distance of $10^3\ell$ and is sampled at fixed spatial increment $\ell$, yielding $n=10^6$ independent observations. Although the distributions of $\epsilon^\prime_{ii}$ for $i=1:3$ are
broad, with standard deviation $\sigma^2_{\epsilon^\prime_{ii}}\sim 1$, the large number $n$ of independent observations reduces the standard error of $\lambda_\infty$ in (\ref{eqn:Lyapunov}) to $\sigma^2_{\epsilon^\prime_{ii}}/\sqrt{n}\sim 10^{-3}$.

\end{document}